\documentstyle[12pt,epsfig]{article}

\newlength{\dinwidth}
\newlength{\dinmargin}
\def\lapproxeq{\lower .7ex\hbox{$\;\stackrel{\textstyle
<}{\sim}\;$}}
\def\gapproxeq{\lower .7ex\hbox{$\;\stackrel{\textstyle
>}{\sim}\;$}}

\def\be{\begin{equation}}
\def\ee{\end{equation}}
\def\bea{\begin{eqnarray}}
\def\eea{\end{eqnarray}}

\begin{document}
\titlepage
\begin{flushright}
IPPP/01/37 \\
DCPT/01/74 \\
CERN-TH/2001-273 \\
Cavendish-HEP-2001/14 \\
October 2001 \\
\end{flushright}

\vspace*{2cm}

\begin{center}
{\Large \bf MRST2001:  partons and $\alpha_S$ from precise deep inelastic} \\

\vspace*{0.4cm}
{\Large \bf scattering and Tevatron jet data} \\

\vspace*{1cm}
A.D. Martin$^a$, R.G. Roberts$^b$, W.J. Stirling$^a$ and
R.S. Thorne$^{c,}$\footnote{Royal Society University Research Fellow} \\

\vspace*{0.5cm}
$^a$ Department of Physics and Institute for
Particle Physics Phenomenology, University of Durham, Durham, DH1
3LE \\

$^b$ Theory Division, CERN, 1211 Geneva 23, Switzerland \\

$^c$ Cavendish Laboratory, University of Cambridge, Madingley
Road, Cambridge, CB3 0HE
\end{center}

\vspace*{2cm}

\begin{abstract}
We use all the available new precise data for deep inelastic and
related hard scattering processes to perform NLO global parton
analyses. These new data allow an improved determination of
partons and, in particular, the inclusion of the recent
measurements of the structure functions at HERA and of the
inclusive jets at the Tevatron help to determine the gluon
distribution and $\alpha_S$ better than ever before. We find a
somewhat smaller gluon at low $x$ than previous determinations and
that $\alpha_S (M_Z^2) = 0.119 \pm 0.002 \,\hbox{(\rm expt.)} \pm
0.003 \,\hbox{(\rm theory)}$.
\end{abstract}

\newpage

\section{Introduction}
Recently a great deal of new data have become available which help
to determine the parton distributions of the proton. In particular
we have new measurements of the structure functions by the H1
\cite{H1A,H1B,H1C} and ZEUS \cite{ZEUS} collaborations at HERA,
and of the inclusive jet distribution by the D0 \cite{D0} and CDF
\cite{CDF} collaborations at the Tevatron. These new data are both
more precise and extend the kinematic range, and thus constrain
the parton distributions, and the strong coupling $\alpha_S$, more
tightly than ever before. ZEUS have also released a new
measurement of the charm contribution to the structure function
\cite{ZEUSc} which, although it still has large errors, covers a
wider kinematic range than previously. In addition, the CCFR
collaboration \cite{CCFR} have re-analysed their neutrino data in
a model independent way and the discrepancy with the NMC data for
$x \lapproxeq 0.1$ has been largely resolved.  Also NuTeV data are
becoming available \cite{NuTeV}, both for single and double muon
production, and are improving the constraints from the neutrino
sector.  The E866 collaboration \cite{E866} also have increased
statistics for $pp$ and $pn$ Drell-Yan production and improve the
determination of the difference between the $\bar{u}$ and
$\bar{d}$ distributions. Finally we note that as well as the usual
data sets used in our previous fits, i.e. BCDMS \cite{BCDMSp} and
SLAC \cite{SLAC} proton data, NMC proton and deuterium data
\cite{NMC}, E665 proton data and deuterium data \cite{E665}, CCFR
data on $F_3^{\nu(\bar \nu)N}(x,Q^2)$ \cite{CCFR3}, E605 Drell-Yan
data \cite{E605} and CDF $W$-asymmetry data \cite{Wasymm} we have
also included BCDMS \cite{BCDMSd} and SLAC deuterium data
\cite{SLAC} in order to obtain as precise a determination of the
separate contributions of the up and down valence quarks at high
$x$ as possible.\footnote{For all deuterium structure functions we correct
for shadowing effects \cite{shaddeut}.}
We also include the most recent ZEUS SVX data \cite{ZEUSSVX} since it spans a 
slightly different range to that in \cite{ZEUS}.   
We no longer include prompt photon data due to
theoretical problems and possible inconsistencies between data
sets, and instead allow the high $x$ gluon to be determined by the
vastly improved Tevatron jet data.

We note that both H1 \cite{H1C} and ZEUS \cite{ZEUS} have recently
performed NLO DGLAP fits to their respective data, supplemented in
the former case by BCDMS data with $y_\mu > 0.3$, and in the
latter case by BCDMS, NMC, E665  and CCFR $F_3^{\nu(\bar
\nu)N}(x,Q^2)$ data. In particular, the H1 analysis determines
$\alpha_S$ and the gluon simultaneously.  A value
\begin{equation}\label{eq:a1}
  \alpha_S (M_Z^2) \; = \; 0.1150 \: \pm \: 0.0017 (\rm expt.)
  \begin{array}{l} + 0.0009 \\ - 0.0005 \end{array} \: ({\rm
  model})
\end{equation}
is obtained, with an additional uncertainty of about $\pm 0.005$,
mainly due to the uncertainty in the renormalization scale. A
preliminary ZEUS analysis, reported at DIS2001, quoted
$\alpha_S(M_Z^2)= 0.1172 \: \pm \: 0.0008 ({\rm uncor.}) \: \pm \:
0.0054 ({\rm cor.})$ \cite{ZEUS01}. 
There is also an analysis including some of both the recent H1 and ZEUS data
along with NMC, SLAC and BCDMS data, and which allows higher twist 
contributions, which obtains $\alpha_S(M_Z^2)= 0.1171 \: \pm \: 0.0015 
({\rm expt.})$ \cite{alekhin}.  
We will find that the
inclusion of additional data sets tend to increase these values
somewhat. As an example of this we emphasize that the Tevatron jet
data are an important ingredient in pinning down the value of
$\alpha_S$ from deep inelastic scattering (DIS) and related data,
since they provide the dominant constraint on the gluon at large
$x$.  In fact, the inclusion of new jet data \cite{D0,CDF} into
the global analysis considerably improves the determination of the
gluon. For example, we find an uncertainty of about 15\% on the
gluon distribution at $x = 0.4$ and $Q^2 = 20~{\rm GeV}^2$, but
this is correlated with the value of $\alpha_S(M_Z^2)$.

\section{The new optimum parton set}

We perform a global NLO DGLAP analysis incorporating all the
high-precision data mentioned in the Introduction. The evolution
begins at $Q_0^2 = 1~{\rm GeV}^2$ where there are three active
quark flavours.  We work in the $\overline{\rm MS}$
renormalization scheme and use the Thorne-Roberts \cite{TR}
variable flavour number procedure to evolve through the charm and
bottom thresholds. We note that we let all data sets in the fit carry 
equal weight. This is because we now feel that the full set of data is 
spread relatively evenly over the kinematic range of $x$ and $Q^2$, and also
over the different partons, e.g. the Tevatron jet data is now extensive enough
to determine the high $x$ gluon accurately, and we no longer have to 
give existing data a high weight to tie down this particular region of 
parton space. In fact those data sets with very few points, e.g. the E866
Drell-Yan asymmetry measurements, probe partons 
(in this case $\bar d -\bar u$) to which 
the rest of the data are rather insensitive, and the few points with unit 
weight are sufficient. We note that our best fit gives a $\chi^2$ per point 
of about 1 for all data sets (except for the E605 Drell-Yan data, 
for the reason given in section 6) justifying the equal weighting.     

As well as deciding on data sets and weights, we have to
decide on a set of cuts in the usual variables $Q^2$, $W^2$ and
$x$. In order to investigate this we made a study of the
sensitivity of the analysis to variation of these data cuts. We
discovered that there was only marginal evidence for an
improvement in quality if the $Q^2$ cut was raised from $2~{\rm
GeV}^2$ to $3~{\rm GeV}^2$ and no marked improvement above this.
There was a marked improvement in quality if $W^2$ is raised from
our previous cut of $10~{\rm GeV}^2$ until we reach $12.5~{\rm
GeV}^2$, which may easily be interpreted as due to the influence
of higher twist and/or large $\ln(1-x)$ terms in the perturbative
expansion. Hence, for the global fit presented below DIS data with
$Q^2 > 2~{\rm GeV}^2$ and $W^2 > 12.5~{\rm GeV}^2$ are included,
in order to exclude regions where higher twist and/or higher
orders in $\alpha_S$ are expected to play an important role. We
also found that if a lower cut in $x$ was introduced there was
continual improvement in the quality of the fit until $x$ reached
a value of about 0.005, suggesting that $\ln(1/x)$ terms in the
perturbative series may be important. The results and consequences
of these cuts, particularly that in $x$, will be dealt with in a
future paper \cite{MRSTcut}, but for the present analysis we take
the conventional approach of not using any $x$ cut and
investigate/suffer the consequences.

The optimum global NLO fit is obtained with the starting
parameterizations of the partons at $Q_0^2 = 1~{\rm GeV}^2$ given
by
\begin{eqnarray}
\label{eq:a2}
 x u_V & = & 0.158 x^{0.25} (1-x)^{3.33}(1+5.61x^{0.5}+55.49x) \\
\label{eq:a3}
 x d_V & = & 0.040 x^{0.27} (1-x)^{3.88}(1+52.73x^{0.5}+30.65x) \\
\label{eq:a4}
 x S & = & 0.222 x^{-0.26} (1-x)^{7.10}(1+3.42x^{0.5}+10.30x)\\
\label{eq:a5}
 x g & = & 1.90 x^{0.09} (1-x)^{3.70}(1+1.26x^{0.5}-1.43x)
-0.21 x^{-0.33} (1-x)^{10}.
\end{eqnarray}
The flavour structure of the light quark sea is taken to be
\begin{equation}\label{eq:a6}
  2\bar{u}, 2\bar{d}, 2\bar{s} \; = \; 0.4S - \Delta, \quad 0.4S +
  \Delta, \quad 0.2S
\end{equation}
with $s = \bar{s}$, as implied by the NuTeV data \cite{NuTeV}, and
where
\begin{equation}\label{eq:a7}
  x \Delta \; = \; x (\bar{d} - \bar{u}) \; = \;
1.195 x^{1.24} (1-x)^{9.10}(1+14.05x-45.52x^2)
\end{equation}
The masses of the quarks are taken to be $m_c =1.43~{\rm GeV}$ and
$m_b = 4.3~{\rm GeV}$, the former giving the best fit to the charm
structure function data. The optimum fit corresponds to
$\alpha_S(M_Z^2) = 0.119$, i.e.\ $\Lambda_{\overline{\rm MS}}(n_f
= 4) = 323~{\rm MeV}$, in very good agreement with the world
average.\footnote{We use the matching between the $n_f$- and 
$n_{f+1}$-flavour couplings calculated in \cite{flav1}, 
and corrected in \cite{flav2},
up to NLO in $\alpha_S$. At this order the coupling is continuous across 
threshold but the derivative is discontinuous. 
More details may be found in section 3 of the first of \cite{TR}.}    
We estimate the error in $\alpha_S(M_Z^2)$ due to the
errors on the data fitted in the global analysis to be about $\pm
0.002$, as will be explained in detail later. The resulting
partons are shown in Fig. 1.

The improved HERA data greatly increase the constraints on the
gluon at small $x$.  The extra term in (\ref{eq:a5}) is required
to achieve an acceptable fit, and allows the starting gluon to
become negative at small $x$.  A fixed $(1-x)^{10}$ behaviour is
incorporated in this extra term so that only the small $x$ form of
the gluon is affected. Not including this additional term, which
allows the input gluon to be negative at small $x$, would lead to
the input gluon being strongly valence-like and to a global
increase in $\chi^2$ of about 100. Note that only half this
increase comes from the data points at very low $x$ (say
$x<0.001$), the rest coming from the HERA and NMC points in the
region $0.001<x<0.1$, as will be discussed in the next section.
The gluon in the present analysis becomes positive for all $x >
10^{-5}$ and $Q^2 > 5~{\rm GeV}^2$, and for $Q^2 > 2-3~{\rm
GeV}^2$ for $x > 10^{-4}$. We note that while a negative gluon 
distribution may be slightly disturbing, there is no real reason for worry 
since the gluon distribution is not a real physical quantity, particularly in 
a somewhat unphysical factorization scheme such as ${\overline {\rm MS}}$. 
The implications for physical quantities will be discussed in the next 
section.

Recall that in the MRST99 analysis \cite{MRST99} the uncertainties
in the gluon were illustrated by presenting the optimal fit $g$
together with two fits $g\!\uparrow$ and $g\!\downarrow$, with
larger and smaller gluons at large $x$, which represented the
extremes of acceptable descriptions of the data.  The present
analysis, with greatly improved data, significantly reduces the
uncertainty in the gluon distribution and yields an optimal
solution with a large $x$ gluon nearer to $g\!\uparrow$ than to
$g$.  For this reason in Fig.~2 we compare the present partons
with those of the $g\!\uparrow$ set of MRST99 \cite{MRST99}.  We
see that the major difference is in the gluon, or is a consequence
of this changed gluon.  First, we note the extended
parameterization for the gluon, required by the new HERA data,
leads to a far smaller gluon at the lowest $x$ and $Q^2$. Since
the quarks are determined by evolution driven by the gluon in this
range, they are also smaller than their MRST99 counterparts. The
gluon is also smaller in the range $x\sim 0.3$ than that for
MRST99$(g\uparrow)$ --- the jet data requiring less gluon in this
range than the prompt photon data with no intrinsic $k_T$
included. Both these reductions in the gluon allow for slightly
more gluon in the range $x\sim 0.1$, giving an increased
$dF_2(x,Q^2)/d\ln Q^2$ for $x$ a little below this. The shape of
the charm (and, to a lesser extent, the strange) distribution
simply follows the gluon since it is generated mainly by evolution
from the gluon. Finally we note that the down quark is slightly
smaller at high $x$ than in MRST99 due to the effect of the extra
deuterium data included in the present fit, and as a consequence it is 
slightly larger for values of $x$ in the region of $0.01$. 

\section{The description of the DIS data}

A good description of the HERA data is obtained, as can be seen
from Figs.~3-6.\footnote{Note that as in previous fits we have
effectively fit the published cross section rather than
$F_2(x,Q^2)$ at the larger values of $Q^2$, i.e. we have corrected
for our own values of $F_L(x,Q^2)$ rather than use those obtained
by the fits by the experiments.} Compared to MRST99, the curves
for $F_2$ are flatter in $Q^2$ for $x<0.001$, but slightly steeper
at higher $x$. In fact for $0.001 < x <0.01$ the data,
particularly the high $Q^2$ H1 and the NMC data, would prefer a
higher $d F_2/d \ln Q^2$, as can be seen in Figs. 5 and 6, where the low 
$Q^2$ NMC data is consistently below the theory and the high $Q^2$ H1 data 
is mostly above the theory. It appears as though the ZEUS data at high $Q^2$ 
tend to lie below the H1 data, and are more consistent with the fit.
However, we note that the preliminary ZEUS 98-99 data for $Q^2 \geq 
200~{\rm GeV}^2$ \cite{ZEUS9899} seem to be more in line with the H1 data 
and prefer a steeper slope (i.e. they lie consistently above MRST99 and hence 
would also be above MRST2001). The
systematic failure in this region of $x$ is a cause for concern regarding
the validity of an NLO fit. In fact we note that the
aforementioned improvement to the fit in this region, which comes
from allowing the negative input gluon at small $x$, is because
this form allows there to be more gluon in the moderate $x$ region
(from the momentum sum rule), and hence a larger value of
$dF_2(x,Q^2)/d\ln Q^2$, as preferred by the data. We also note
that although at the lowest $x$ and $Q^2$ our gluon distribution
is considerably negative, $d F_2/d \ln Q^2$ is quite clearly
positive. This highlights the fact that the frequently quoted
relationship $d F_2/d \ln Q^2 \propto \alpha_S(Q^2) xg(x,Q^2)$ is
not even approximately true at small $x$ and $Q^2$ when one works
beyond leading order in perturbative QCD. Qualitative arguments
about evidence for saturation {\it etc.} which rely on this
relationship should be treated with caution.

The fits to HERA data have been performed using the simplistic
procedure of adding the systematic and statistical errors in
quadrature. However, we have actually performed an analysis of the effects
of the correlated errors. To be specific we have first performed fits 
to the data with
only uncorrelated errors, then let the contributions of the
correlated errors come into effect, and finally iterated. We
find that the absolute value of the $\chi^2$ using this procedure
increases for the ZEUS data, and stays more or less constant for
the H1 data, but that somewhat surprisingly the position of the minima 
and incremental changes in $\chi^2$ when
comparing different theoretical results stays much the same as
when using the more simplistic addition in quadrature of all
errors. Hence, we decide to present the simpler procedure in
determining the partons, and discuss details of the effects of correlated
errors in an Appendix since the results
with full errors turn out to be an unnecessary complication. 
We do however let the ZEUS normalization go to its lower limit of $98\%$ 
in our fits in order to obtain the optimum description.

The comparison with the charm data \cite{ZEUSc,H1charm} can be
seen\footnote{Updated charm data from the H1 collaboration have
recently become available \cite{H1NC}, but the results depend on
which Monte Carlo is used to extrapolate over the full range of
phase space.  Since these data are similar to their previous charm
data \cite{H1charm}, we show only the latter in Fig.~7.} in Fig.
7. As one can see it is of a perfectly acceptable quality, and the
errors on this data are still large. There is, however, a slight
tendency to undershoot the data at the lowest values of $x$ and
$Q^2$, and this may be a sign of the need to improve the
theoretical treatment in this region. In this region of low $x$ and $Q^2$ 
the prediction for the charm structure function is a little smaller than 
that for the MRST99 partons, which is entirely due to the smaller gluon we
now find in this region. Finally we note that the fit to the higher $x$ 
EMC charm data \cite{EMCcharm} is very similar to that for the default set 
in Fig. 27 of \cite{MRST98} (the high $x$ gluon is now a little larger, but
$m_c$ is $1.43~{\rm GeV}$ rather than $1.35~{\rm GeV}$), 
and hence is perfectly acceptable.  

The prediction for $F_L(x,Q^2)$ is shown in Fig.~8, which also
shows the prediction of the MRST99 partons. We see
that the increased precision of the HERA data and the increased
flexibility of the gluon parameterization have led to a
significant decrease in the prediction for $F_L(x,Q^2)$ at low $x$
and $Q^2$, replaced by a slight increase for $x \sim 0.05$.
Indeed, it now seems as though $F_L(x,Q^2)$ is taking a distinctly
unphysical form for low $x$ and $Q^2< 5~{\rm GeV}^2$, and for part
of this range is negative, and therefore certainly
disallowed.\footnote{In principle, it is internally inconsistent
to fit to $F_2 (x, Q^2)$ data in a region where the predicted
values of $F_L (x, Q^2)$ are negative, namely $x < 10^{-4}$ and
$Q^2 \gapproxeq 2~{\rm GeV}^2$. Since only 6 points are affected,
carrying practically no weight in the fit, we do not remove these
points.} This is a direct consequence of the negative nature of the 
gluon distribution at small $x$ and $Q^2$, and may be taken as another clear 
sign that the standard NLO fit is not working completely properly at small
$x$.\footnote{We do not compare to the H1 extraction of
$F_L(x,Q^2)$ \cite{H1FL} since the different assumptions used in
our fit lead to significantly different forms for the gluon and
for $\alpha_S(M_Z^2)$, and hence different extrapolations into the
high $y$ region.} As far as we are aware $F_L(x,Q^2)$ is the most direct 
probe of the gluon distribution at small $x$ and $Q^2$, and is the most 
appropriate quantity to examine in order to see the real pathological 
effects of the negative gluon distribution. $F_2^c(x,Q^2)$ is less sensitive 
since at low $Q^2$ the kinematic constraint on charm production 
($W^2 \geq 4m_c^2$) means one is probing the gluon at higher $x$ than for
$F_L(x,Q^2)$, and as seen in Fig. 7, $F_2^c(x,Q^2)$ is perfectly well behaved 
down to $x<0.0001$. 

At higher $x$ the main change in our fit is due to the reanalysis
of the CCFR data \cite{CCFR}. Their reanalysis no longer extracts
$F_2(x,Q^2)$ by modeling both $F_L(x,Q^2)$ and $\Delta
xF_3(x,Q^2)$, but now extracts either  $F_L(x,Q^2)$ or $\Delta
xF_3(x,Q^2)$ (there is a high degree of correlation between these)
and $F_2(x,Q^2)$ separately in a Physics Model Independent manner.
This has gone a long way towards resolving the apparent
discrepancy between CCFR and NMC data on $F_2(x,Q^2)$, where it
had previously been impossible to simultaneously fit both for
$x<0.1$. The quality of the fit to the new CCFR $F_2(x,Q^2)$ data
is shown in Fig. 9. Overall the fit is very good. One might argue
that there is still a systematic problem at the lowest $Q^2$, but
this is far less pronounced than with previous analyses. 
There is also potentially a small error associated with the shadowing 
corrections (details of which are found in \cite{MRST98}) 
which we do not account for. The
reanalysis has also established the validity of the previous
$F_3(x,Q^2)$ neutrino data. These data are essentially unaffected
by the reanalysis, but we are now confident in using the data over
the whole $x$ range, rather than just for $x\geq 0.1$. The fit is
good over the whole range of $x$. Note that we normalize the
complete set of CCFR data up by $1\%$ in order to obtain the best
fit.

Other than this, the other new DIS data (at least new for our fit)
at high $x$ are simply the SLAC and BCDMS deuterium data, which we
have introduced for the first time. The fit to these data is shown
in Fig. 10. It is of a perfectly acceptable quality, and one can
see that, as with the proton data, the SLAC deuterium measurements
prefer a rather steeper fall with $Q^2$ than the BCDMS data, and
consequently a larger $\alpha_S$. The SLAC data are normalized up
by $2.5\%$ and the BCDMS down by $2\%$.

\section{Tevatron Jet Data and the Gluon}

One of the major differences between the MRST2001 partons and our
previous parton sets is the manner in which the high $E_T$
Tevatron jets have been included. In the past \cite{MRST98} we
have simply checked that there is reasonable agreement with our
predictions and the jet data. The difficulties in using the prompt
photon data in order to determine the high $x$ gluon combined with
the considerable improvement in Tevatron jet data \cite{D0,CDF}
has led to a change in emphasis. Besides the increase in
precision, the D0 jet data are available in a range of rapidity
intervals and so constrain the partons, and the gluon in
particular, over a much wider $x$ range.  We, therefore, now
include the D0 and CDF jet data in the global fit on an equal
footing with all other data sets.\footnote{The D0 collaboration have recently 
produced data using the $k_T$ algorithm rather than the more usual cone 
algorithm \cite{D0kt}. The agreement between the two methods is moderate,
the major difference being at low $E_T$. We use the original data since 
we feel these have been more extensively studied, and because they 
also cover a much 
wider range in pseudo-rapidity.} However, because in this case
the correlated systematic errors are the dominant source of error,
being much larger than the uncorrelated errors, it is imperative
to deal with these in a correct manner. In fact we adopt the same
method of fitting to the data as do the respective experimental
collaborations when describing their own data. Note, however, that
rather than using some NLO prescription such as JETRAD
\cite{JETRAD} or EKS \cite{EKS} to generate an NLO correction for
each point, we derive a smooth NLO K-factor by fitting to a set of
such points. Since the NLO corrections generated from the above
programs have some error and scatter, this means that our value of
$\chi^2$ will not be identical to that obtained by the experiments
themselves for the same parton set (though it will be very close).
In particular our values of $\chi^2$ for the CDF1B jet data are a
little higher, while for the D0 jet data they are slightly lower.

We find that for our best global fit we obtain a reasonable description of
the combined jet data with a $\chi^2$ of 170 for 113 points. The
quality of the fits is shown in Figs. 11 and 12 --- the error bars
account for uncorrelated errors alone. In both cases it is clear
that while at the low $E_T$ end the normalization is about
correct, at higher $E_T$ (and rapidity) the theory lies below the
data. An acceptable fit is then obtained by accounting for the
correlated systematic errors. 

First, for the CDF1B fit the
(data$-$theory) obtained is allowed to move by letting data move
relative to theory by application of each of the sources of
correlated error, i.e. the $\chi^2$ is obtained from 
\begin{equation}\label{eq:sys1}
  \chi^2=\sum_i\frac{(T_i/F_i-E_i)^2}{(\Delta E_i)^2} + \sum_k s_k^2
\end{equation}
with
\begin{equation}\label{eq:sys2}
  F_i = 1 + \sum_k f_i ^k s_k,
\end{equation}
where $T_i$ is the theory prediction for data point $i$, $E_i$ is the 
measurement with uncorrelated error $\Delta E_i$, and $f_i^k$ is the 
one-sigma correlated systematic error for point $i$ from error source $k$.
Hence, the data and theory move relative to each other 
at the cost of an increase in $\chi^2$ of $\sum_k s_k^2$,
where $s_k$ is the fraction of one-sigma which has been utilized for each
error source.\footnote{Since these correlated errors are
expected to cut off rather more sharply than Gaussian errors we
limit each $s_k$ to 1. This does not affect any results at all
significantly.} 
Hence in Fig. 12 we see that the effect of introducing the error 
correlations $F_i$ has been to bring down the 
data at higher $E_T$ significantly in order to match the shape of the
theory prediction. This large shift requires many of the $s_k$ to be 
of the order of 1, and for them all to conspire to move the data in the same 
direction relative to the theory. 

For the D0 jet data the fit is performed 
using the full error matrix, i.e. 
\begin{equation}\label{eq:sys3}
  \chi^2=\sum_{i,j}(E_i-T_i)[(T_i/E_i)C_{ij}(T_j/E_j)]^{-1}(E_j-T_j),
\end{equation}
where $C_{ij}$ are the covariance matrix elements defined by
\begin{equation}\label{eq:sys4}
  C_{ij}=\sum_k \rho^k_{ij} \Delta E^k_i \Delta E^k_j,
\end{equation}
where $k$ runs over all sources of error, $\Delta E_i^k$ is the error of 
point $i$ and $\rho^k_{ij}$ is the correlation between points $i$ and $j$.
This is actually a very similar way to obtain a $\chi^2$ to the previous 
method (see Appendix A of \cite{CDF}). However,  
it accounts for the correlated systematic
errors in a rather less transparent manner, though the good fit to the D0 
jet data must clearly be obtained in much the same way. We illustrate the 
correlated errors in Fig. 11 simply by introducing a band with width given 
by adding each source of correlated error in quadrature. Although this is 
not as explicit as in Fig. 12, it indicates roughly how the data may move 
relative to the theory without a large cost in $\chi^2$.    

At the central value of $\alpha_S(M_Z^2)$ of 0.119, this global
fit (including jet data) allows a variation in $g(x,Q^2)$ of about
$5\%$ for $x>0.1$ and $Q^2 \sim 2000~{\rm GeV}^2$, which
corresponds to $10-15\%$ accuracy for $x>0.1$ and $Q^2 \sim
20~{\rm GeV}^2$. This is a factor of 3 or so less than the MRST99
variation of the gluon ranging from the $g\!\uparrow$ to the
$g\!\downarrow$ gluon, and hence we do not provide parton sets
with gluon extremes. Since the body of jet cross section data is
(very roughly) $\propto \alpha_S(E_T^2/4)g(x,E_T/2)$, then
$g(x,\mu^2)$ for $\mu^2$ of order $10^3~{\rm GeV}^2$ is roughly
inversely proportional to $\alpha_S(M_Z^2)$. However, at high $x$
the gluon distribution decreases more rapidly with increasing
$Q^2$, the larger the value of the coupling. This increase in
speed of evolution with increasing $\alpha_S$ more than
compensates for the decrease in the high-scale gluon required by
the jets with $\alpha_S$, and for low $Q^2$ ($Q^2 \sim 10~{\rm
GeV}^2$) the high $x$ gluon increases as $\alpha_S (M_Z^2)$
increases.

We note, however, that these optimum global fits are not the best
possible fits to the high $E_T$ jet data. The fit is only achieved
by compensating for the smallness of the theory at high jet $E_T$
and $\eta$ (both of which probe the highest $x$) using the
correlated systematic errors. Hence, the fit can be improved by an
increase in the size of the high $x$ gluon. In principle it is
possible to obtain a fit with a $\chi^2$ of about 120 for the 113
points (see below) rather than $\chi^2=170$, the scatter of data
points making this $\chi^2$ value about the lowest that is
achievable. At the central $\alpha_S(M_Z^2)$ value of 0.119 it is
possible to raise the high $x$ gluon sufficiently to improve the
quality of the jet fit to $\chi^2=135$, but only at the cost of
$\Delta \chi^2=60$ for the rest of the data. This is mainly at the
expense of the description of the moderate $x$ DIS data, i.e. H1,
ZEUS and NMC data, since the increase of gluon at high $x$ is
countered by a decrease at intermediate $x$, and hence a decrease
in $dF_2(x,Q^2)/d\ln Q^2$. At lower $\alpha_S$ the price is even
higher since the lower $\alpha_S$ already impacts upon the
behaviour of $dF_2(x,Q^2)/d\ln Q^2$.

However, as one goes to higher $\alpha_S(M_Z^2)$ the situation
changes. At $\alpha_S(M_Z^2)=0.121$ one can obtain a fit to the
jet data with $\chi^2=118$, and where this improvement is only
marginally overcompensated by the deterioration in the rest of the
fit compared to the best global fit. For this set of partons,
denoted by MRST2001J or simply J, the fit to the jet data is shown
in Figs. 13 and 14. For the fit to the D0 data the shape is
obviously greatly improved, both as a function of $E_T$ and of
$\eta$, demonstrating that the apparent excess $\chi^2$ in the
description of the Tevatron jet data is either a problem of parton
distributions or of systematic errors, but is unlikely to be a
sign of new physics. The normalization of the theory is a little
high, but this is easily accounted for by the systematic error in
normalization. The fit to the CDF1B data actually gives a slightly
worse $\chi^2$ than before. But now the fit does not rely on a
large shift of data due to systematic errors, and is perhaps more
satisfactory in this sense. The problem with this J set of partons
is the behaviour of the gluon.  The input form at $Q_0^2 = 1~{\rm
GeV}^2$ is
\begin{equation}\label{eq:jetglu}
 x g  =  123.5 x^{1.16} (1-x)^{4.69}(1-3.57x^{0.5}+3.41x)
-0.038 x^{-0.5} (1-x)^{10},
\end{equation}
which is shown in Fig.~15, together with the behaviour at $Q^2 =
20~{\rm GeV}^2$.  We see that the input shape has a rather
worrying ``kink'' which results in the distinct ``shoulder'' at
$Q^2 = 20~{\rm GeV}^2$.\footnote{This seems to be made possible by the 
interplay between a very large coefficient for the first term in 
(\ref{eq:jetglu}),
the large power of $x$, i.e. $x^{1.16}$ in this term, and the extra effect of 
the second term controlling the very small $x$ behaviour. This second term 
then effectively frees one parameter in the first term, which for previous 
parton sets represented the full parameterization of the input gluon,
allowing more flexibility in the high $x$ form of the gluon.} 
We do not deem this to be an acceptable
gluon (admittedly a subjective decision), and rule this fit out,
although we do make the MRST2001J set of partons available. At
$\alpha_S(M_Z^2)=0.120$ we obtain a similar result, i.e. the best
overall fit gives about $\chi^2=135$ for the jet data, but has a
gluon with the same type of peculiarities (though less severe).
Again we rule this fit out. Hence, in our fits we impose the
condition that $d^2(xg(x,Q_0^2))/dx^2$ does not change sign in the
region of high $x$ which rules out the possibility of both kinks
and shoulders in the high $x$ gluon distribution. Imposing this
condition results in a fit to the jet data within the global fit
which is roughly independent of $\alpha_S(M_Z^2)$.\footnote{We
note that a very good fit to the jet data could be achieved for
$\alpha_S(M_Z^2) < 0.118$ with a gluon without peculiarities, but
that this results in a fit to the rest of the data which is very
poor. This problem is improved, though not completely rectified,
when an $x$-cut is applied (see \cite{MRSTcut}).}

\section{Description of other data}

The fit to much of the rest of the data is very much along the
lines discussed in detail in \cite{MRST98}. The NuTev data
\cite{NuTeV} on single and double muon production do not
qualitatively change the conclusions regarding the strange
contribution to the sea already indicated by the CCFR dimuon data
\cite{CCFRdimuon}, i.e. that the strange distribution is
acceptably obtained from half the average of the $\bar u$ and $\bar
d$ distributions at $Q_0^2 = 1~{\rm GeV}^2$. Similarly the E866
collaboration \cite{E866} have provided new data on the Drell-Yan
asymmetry which is more accurate and extends the kinematic range
slightly, but does not really change the relative $\bar{u},
\bar{d}$ behaviour of the partons. In particular, our simple
parameterization of $\bar d -\bar u$ still suggests that $(\bar
d/\bar u) \leq 1$ for $x>0.35$, but there is no evidence whether
this is really true or not. The lepton rapidity asymmetry data
from CDF \cite{Wasymm} (related to the $W$ rapidity asymmetry)
also continue to give us important information on the $u/d$ ratio.

Finally the E605 Drell-Yan data \cite{E605} still play an
important role in pinning down the form of the sea quarks at high
$x$. However, they also play an important and unexpected role in
influencing the fit to the jet data and determining
$\alpha_S(M_Z^2)$. As we will see below, the quality of the
description of these data deteriorates as $\alpha_S(M_Z^2)$
increases. This is actually an indirect effect. As
$\alpha_S(M_Z^2)$ increases, the high $x$ gluon at lower $Q^2$
increases so as to give the correct gluon normalization when
evolved up to the scales appropriate for the description of the
jet data. This larger high $x$ gluon (and larger
$\alpha_S(M_Z^2)$) drives a positive evolution of the high $x$ sea
quarks. As this effect becomes more significant it distorts the
shape of the sea quark distribution in the range relevant for
fitting Drell-Yan data, worsening the fit. Therefore the E605 data
prefer lower value of $\alpha_S(M_Z^2)$ and a lower high $x$
gluon. Indeed, for the MRST2001J type partons the fit to the
Drell-Yan data deteriorates quite seriously compared to the best
global fit. Hence, these data have assumed a more important role
in the context of the whole global fit than previously.

\section{Quality of Fit and Determination of $\alpha_S(M_Z^2)$.}

The quality of the central fit for the major data sets is shown in
Table 1 below. For each of the smaller data sets, e.g. 
CDF $W$-asymmetry \cite{Wasymm} and E866 Drell-Yan asymmetry \cite{E866}, 
the $\chi^2$ per
degree of freedom is about 1 per point. For all the DIS data sets
the numbers are quoted for statistical and systematic errors added
in quadrature. The quality of the fits to the individual data sets
is satisfactory.  For the E605 data the systematic errors are
quoted in a slightly ambiguous manner, and are generally
subdominant, and so we fit to statistical errors alone. Hence, the
quite large $\chi^2$ in this case. The treatment of the correlated
systematic errors for the Tevatron jet data has been discussed in
Section~4.

\begin{table}[h]
\caption{Quality of the fit for MRST2001 partons to different data
sets.  The first MRST column shows the $\chi^2$ values of the
optimum fit with $\alpha_S (M_Z^2) = 0.119$.  Also shown are the
values for parton sets obtained from fits with $\alpha_S (M_Z^2) =
0.117$ and $0.121$, as well as those for parton set J which has
structure in the high $x$ gluon.}
\begin{center}
\begin{tabular}{|cccccc|}\hline
Data set          & No. of   & MRST & MRST & MRST & MRST  \\
                  & data pts &      & 0.117 & 0.121 & J     \\ \hline
H1 $ep$           & 400      & 382  & 386  & 378  & 377   \\
ZEUS $ep$         & 272      & 254  & 255  & 258  & 253   \\
BCDMS $\mu p$     & 167      & 193  & 182  & 208  & 183   \\
BCDMS $\mu d$     & 155      & 218  & 211  & 226  & 219   \\
NMC $\mu p$       & 126      & 134  & 143  & 127  & 135   \\
NMC $\mu d$       & 126      & 100  & 108  & 95   & 100   \\
SLAC $ep$         & 53       & 66   & 71   & 63   & 67    \\
SLAC $ed$         & 54       & 56   & 67   & 47   & 58    \\
E665 $\mu p$      & 53       & 51   & 50   & 52   & 51    \\
E665 $\mu d$      & 53       & 61   & 61   & 61   & 61    \\
CCFR $F_2^{\nu N}$& 74       & 85   & 88   & 82   & 89    \\
CCFR $F_3^{\nu N}$& 105      & 107  & 103  & 112  & 110   \\
NMC \it{n/p}      & 156      & 155  & 155  & 153  & 161   \\
E605 DY           & 136      & 232  & 229  & 247  & 273   \\
Tevatron Jets     & 113      & 170  & 168  & 167  & 118   \\
\hline Total      & 2097     & 2328 & 2346 & 2345 & 2337  \\
\hline
\end{tabular}
\end{center}
\end{table}

The way in which the quality of the fit to both the total and to
each data set varies with $\alpha_S(M_Z^2)$ is shown in detail in
Fig. 16. It must be remembered that the quality of the fit for a
single data set within the context of a global fit is not the same
thing as the quality of the fit for that set alone, and many sets
influence each other strongly. Nevertheless, one can pick out some
interesting facts from Fig. 16.

For the DIS data sets it is clear that only the two BCDMS sets
strongly prefer lower values of $\alpha_S(M_Z^2)$. These are more
than compensated by the SLAC and NMC data sets, which for both
proton and deuterium structure functions strongly prefer higher
values of $\alpha_S(M_Z^2)$. Both CCFR data sets are relatively
insensitive to the value of the coupling, at least for $0.116 <
\alpha_S(M_Z^2) < 0.122$. This also appears to be true for the H1
and ZEUS data sets. However, this latter apparent insensitivity is
due to the fact that the combined HERA data sets carry a lot of
weight in the fit, and the gluon distribution at small $x$ is
largely determined by ensuring that these data are fit well. This
is therefore just a manifestation of the long-established fact
that the small $x$ gluon and the value of $\alpha_S(M_Z^2)$ are
completely correlated in fits to the HERA data, and without any
additional handle on the gluon\footnote{The charm structure
function is strongly correlated to the evolution of the total
structure function, and therefore does not provide an independent
constraint.} there is no way to remove this. It is clear from Fig.
16 that if one takes only BCDMS data as well as HERA data, as in
the H1 analysis \cite{H1C}, one will determine a low value of
$\alpha_S(M_Z^2)$, but taking SLAC or NMC as the additional set a
very different conclusion will be reached.

The combined Tevatron jet data behaves similarly to the HERA data,
i.e. the gluon conspires with $\alpha_S(M_Z^2)$ to give roughly
the same $\chi^2$ for all $\alpha_S$. Interestingly the jet data
and HERA data manage to conspire with each other so that the sum
of their $\chi^2$ remains roughly constant. As $\alpha_S(M_Z^2)$
increases the gluon at moderate $Q^2$ and high $x$ increases to
maintain the fit to the jet data. From the momentum sum rule this
leaves less gluon at small $x$, but the larger $\alpha_S(M_Z^2)$
manages to keep the value of $dF_2(x,Q^2)/d \ln Q^2$ acceptable.
The fact that this trade-off between HERA $F_2(x,Q^2)$ data and
Tevatron jet data results in almost complete insensitivity to
$\alpha_S(M_Z^2)$ strikes us as remarkable. Note, however, that
the constant total $\chi^2$ for the jets is made up of a
contribution from the CDF1B data which increases sharply with
increasing $\alpha_S(M_Z^2)$, and a rapidly falling contribution
from the D0 jet data. In detail one finds that the general
normalization and shape of the theory compared to data improves
with increasing $\alpha_S(M_Z^2)$. This leads to the improvement
in the fit to D0 data. However, the precise shape of the CDF1B
data seems easiest to achieve by obtaining a poor comparison
between theory and data which is then compensated for by quite
large movements coming from the correlated errors. When the shape
and size is nearly correct to begin with this seems to leave less
room for maneuver for the correlated errors to produce exactly
the correct shape (note that the $\chi^2$ is better in Fig. 12
than in Fig. 14). Hence, if one is uncomfortable about letting the
correlated errors conspire to move the data by a large amount the
high $\alpha_S(M_Z^2)$ fits are better.

Finally, as discussed in the last section, the E605 Drell-Yan data
prefer a low value of $\alpha_S(M_Z^2)$. However, this is mainly
due to the correlation between the value of $\alpha_S(M_Z^2)$ and
the high $x$ gluon brought about by the jet and HERA data.

Putting all the contributions together we obtain a total $\chi^2$
which has quite a sharp minimum at $\alpha_S(M_Z^2)=0.119$. We
then adjudge the error in this best value of $\alpha_S(M_Z^2)$,
within the context of an NLO-in-$\alpha_S$ fit, by letting the
$\chi^2$ increase by about 20 units. Clearly it is inappropriate
to base the error on the increase of a single unit, for a variety
of reasons. First, the treatment of the errors in this analysis is
far from statistically rigorous, and even if it were, the errors
themselves are far from having a true Gaussian distribution. Also,
we have made many decisions in performing this analysis, such as
data cuts, the choice of parameterizations of partons, etc.
Changing any of these and refitting would lead to changes in
$\chi^2$ of about $5-10$ for the remaining data, and so our
increase in $\chi^2$ should be at least this value . Making our
choice of an increase of 20 we obtain $\alpha_S(M_Z^2)=0.119 \pm
0.002 ({\rm expt.})$. We see from Fig. 16 that beyond these limits
the global $\chi^2$ increases very quickly.\footnote{We have also
investigated the fits without the Tevatron jet data included. Even
though removing this constraint allows gluons to migrate to lower
$x$ and in principle fit the HERA data with lower $\alpha_S$, the
overall impact on the global fit is not large. The minimum moves
down by $\Delta \alpha_S (M_Z^2) \sim 0.0002$. For rather low
values of $\alpha_S(M_Z^2)$, e.g.\ 0.116 or lower, the removal of
the high $x$ gluon constraint does allow an improvement in the fit
to HERA and NMC data, but at this value of $\alpha_S(M_Z^2)$ the
global fit has become much worse anyway, and all one would obtain
with jet data removed would be a shape for the total $\chi^2$ like
that in Fig. 16, but with the slope on the left-hand side a little
more shallow.}

Thus, we present our determination of $\alpha_S(M_Z^2)$ as
\begin{equation}\label{eq:a11}
  \alpha_S (M_Z^2) \; = \; 0.1190 \: \pm \: 0.002 ({\rm expt.}) \: 
\pm \: 0.003({\rm
  theory}).
\end{equation}
We do not adopt the traditional, but {\it ad hoc}, manner of
obtaining the theoretical error by varying renormalization and
factorization scales up and down by factors of 2 (or of 4). This
takes no account of the errors attributable to higher order
logarithmic enhancements.  For example, in DIS there are additional
logarithms in $(1-x)$ and $1/x$ in the coefficient functions and
splitting functions at higher orders in $\alpha_S$ which
variations in scale tell us nothing about. Similar logarithmic
enhancements also exist for the other quantities fitted, such as
data near threshold. Hence, we obtain our theory error by
comparing with alternative theoretical treatments which do tell us
something more concrete about the missing corrections, i.e.
approximate NNLO fits e.g. \cite{MRSTNNLO}, or fits which attempt
a resummation of $\ln(1/x)$ and $\ln(1-x)$ terms \cite{resum}.
These suggest that $0.003$ is an appropriate theoretical
error.\footnote{These investigations suggest that
$\alpha_S(M_Z^2)$ might move down slightly from 0.119.}

As regards the errors on the partons themselves, in a separate
study we will present the uncertainties in the predictions of key
observables, and show how they reflect the uncertainties on the
parton distributions. An example of this is seen in
\cite{Bologna}. However, as in the case of $\alpha_S(M_Z^2)$, we
believe the theoretical errors to be generally more important than
the experimental errors, particularly in some regions of parameter
space.

\section{Conclusions}

In this paper we have performed global analyses of all the most
up-to-date data on deep inelastic scattering and related processes
in order to best determine the parton distributions and the value
of $\alpha_S(M_Z^2)$ within the context of a conventional NLO fit.
This is an improvement on our previous analyses mainly because of
some very important new sets of data. In particular the new HERA
data \cite{H1A,H1B,H1C,ZEUS} are far more precise than previously
and cover an extended range in $x$ and $Q^2$. Also the new D0 and
CDF Tevatron jet data are again more precise, with systematic
errors which are better understood, and which extend their
previous kinematic ranges. These new HERA and Tevatron data sets
together impose far more stringent limits on the parton
distributions than ever before. We also obtain a tight constraint
on the value of $\alpha_S(M_Z^2)$. Our best overall fit
corresponds to $\alpha_S(M_Z^2)=0.119$, and investigating
variations about this minimum we obtain $\alpha_S (M_Z^2) = 0.119
\pm 0.002 ({\rm expt.}) \pm 0.003({\rm theory})$. The quality of
the fits for the two experimental limits of $0.117$ and $0.121$
can be seen in Fig. 16, and are also detailed in Table 1.

The new data sets have a particularly strong impact on the gluon
distribution. In order to fit the new HERA data well we have been
forced into an extension of our previous input gluon
parameterization, allowing it to become negative at small $x$.
Indeed at $Q_0^2=1~{\rm GeV}^2$ it behaves like $xg(x,Q_0^2) \sim
0.2x^{-0.33}$ for $x<0.001$, and this is necessary not only to
obtain a good fit at low $x$ and $Q^2$, but also to allow enough
gluon at higher $x$ to obtain large enough $(dF_2/d\ln Q^2)$ for
$x \sim 0.01$, and enough gluon at large $x$ for the Tevatron jets. 
As $\alpha_S(M_Z^2)$ increases the very small $x$ gluon becomes more negative.
This is due to a combination of factors, i.e. the change of gluon needed by
the jets at high $x$ and by $(dF_2/d\ln Q^2)$ at medium $x$, but the most 
obvious explanation is that as $\alpha_S(M_Z^2)$ increases the positive effect 
of the quark-gluon splitting function at very small $x$ increases
(particularly the NLO contribution), and the gluon in this region 
correspondingly decreases.    
The result of a negative gluon at low $Q^2$ and
$x$ has been confirmed by backwards evolution in \cite{ZEUS01}
(and to a lesser extent in \cite{H1C}). It will be interesting to
see whether a similar conclusion is obtained by other analyses
\cite{CTEQ5,GRV}. We anticipate that the evolution from positive
definite parton distributions at very low scales \cite{GRV} will
be very difficult to sustain.

The Tevatron jet data constrain the high $x$ gluon (though from
the momentum sum rule and convolutions performed in evolution
equations it also affects lower $x$). These provide a far better
constraint than any previous data, and from the best global fit we
now estimate the uncertainty in the gluon distribution for $x>0.1$
and $Q^2 \sim 20~{\rm GeV}^2$ to be $10-15\%$, with the error
decreasing with increasing $Q^2$. This removes the need to produce
the sets of parton distributions with extreme gluons that were
presented in the MRST98 \cite{MRST98} and MRST99 \cite{MRST99}
analyses. There is, however, the caveat that if we were to allow
the input gluon to have a rather unusual shape and also let
$\alpha_S(M_Z^2)$ be $\geq 0.120$ we can produce a fit which is
much better for the jet data, without too great an expense in
$\chi^2$ for the rest of the data. The best possible jet fit is
obtained for $\alpha_S(M_Z^2)=0.121$, and the parton set is denoted
by MRST2001J. The gluon is shown in Fig. 15, and the quality of
the fit using this set of partons is shown in Table 1. We see that
the total $\chi^2$ is not much higher than the central fit, and is
better than the nominal fit for $\alpha_S(M_Z^2)=0.121$. In
particular the large high $x$ gluon helps to counter the
deterioration in the fit to BCDMS data for increasing coupling,
but has a poor effect on the Drell-Yan E605 data (as discussed in
Section 5). However, we reject this as an acceptable set because
of the structure in the form of the high $x$ gluon at low scales,
but make it available as an alternative set. It is in some senses
similar to the CTEQHJ parton set \cite{CTEQHJ} obtained by forcing
the best fits to previous high $E_T$ jet data, but does not seem
to have quite the same features and moreover, we find that our
very good fit to jet data can only be achieved without a huge cost
in $\chi^2$ to the fit to other data for $\alpha_S(M_Z^2) >
0.120$.

Other than the gluon, and the heavy quark distributions which are
generated entirely by evolution and mainly from the gluon, there
are no really dramatic changes in our parton distributions, as can
be seen in Fig. 2. The inclusion of more deuterium data has caused
a slight decrease in the high $x$ down quark distribution, and a
corresponding increase for $x \sim 10^{-2}$, but there is nothing
else too significant. Indeed, our central value of
$\alpha_S(M_Z^2)$ has not changed much either. Though the changes
are small, it is important to quantify their influence on the precision 
predictions of the
$W$ and $Z$ production cross sections at the LHC and the Tevatron.
Table~2 shows the changes in the predictions for these cross
sections when going from the default MRST99 parton set
\cite{MRST99} to the present set where, for ease of comparison, we
have kept the electroweak parameters unchanged.  We see that the
predicted cross sections have increased by about 1\%. This is partly 
caused by the increase in the down quark distribution in the
relevant $(x, Q^2)$ range, see the second plot in Fig.~2.  The
uncertainty of such predictions and the influence on the parton
uncertainty will be the subject of a future paper.

\begin{table}
\caption{Predictions in $nb$ for $W$ and $Z$ production at the Tevatron
and LHC, compared with those of MRST99 \cite{MRST99}.}
\begin{center}
\begin{tabular}{|c|c|c|c|} \hline
&   & MRST99 & MRST2001 \\ \hline
Tevatron & $B_{\ell \nu} \cdot \sigma_W$ & 2.45 & 2.48 \\
& $B_{\ell^+ \ell^-} \cdot \sigma_Z$ & 0.226 & 0.228 \\ \hline
LHC & $B_{\ell \nu} \cdot \sigma_W$ & 20.3 & 20.5 \\
& $B_{\ell^+ \ell^-} \cdot \sigma_Z$ & 1.87 & 1.89 \\ \hline
\end{tabular}
\end{center}
\end{table}

As a final point we note that the overall quality of the
NLO-in-$\alpha_S$ fit remains fairly good. The raised cut in
$W^2$, from $10~{\rm GeV}^2$ to $12.5~{\rm GeV}^2$ has removed
some deficiencies in the high $x$ slope which may be due to higher
twist or higher orders in $\alpha_S(Q^2)$. Also, it is noticed
that investigating cuts in $Q^2$ implies little evidence for
higher twist at general $x$. However, some of the previous areas
of concern have been increased rather than reduced. It is a
worrying point that the minimum $\chi^2$ values for many
individual data sets within the global fit (Fig. 16) lie outside
the range $\alpha_S(M_Z^2)=0.116-0.122$.  Also, as in previous
fits, there is also still a struggle to get a steep enough
evolution of $F_2(x,Q^2)$ in the region $x\sim 0.01$ as is seen in
Figs. 5 and 6.  Moreover, it is also difficult to obtain enough
high $x$ gluon to get a very good fit to the jet data. These two
points, coupled with the rather slow evolution of $F_2(x,Q^2)$ at
the lowest $x$, combine to produce a gluon which has gone from
being valence-like to very negative at the input scale $Q_0^2 =
1~{\rm GeV}^2$. While this is not necessarily a problem in itself,
it has resulted in a prediction for $F_L(x,Q^2)$ that is
worryingly small at very small $x$ (Fig. 8). Hence there are
implications of problems at small $x$. We have not really
considered the effect of a lower $x$ cut in this paper, but will
demonstrate in a future paper \cite{MRSTcut} that investigating
fits with low $x$ data cut out does have a serious effect on the
partons and has strong implications on the real success of the
standard NLO-in-$\alpha_S$ fit at low $x$. Remember that in the
same way that the high $x$ form of the gluon imposed by jet data
influences small $x$ via the sum rule and convolutions, cutting
out small $x$ data can influence the fit and partons at higher
$x$. However, the effect of varying the $x$ cut on the value of
$\alpha_S$ is minimal, since the major constraint comes from the
evolution of the high $x$ partons.  It is, therefore, not
surprising that the value remains well within our quoted
experimental error, $0.119 \pm 0.002$. On a related point we have
already noticed that extending the theory to (an approximate)
NNLO-in-$\alpha_S$ does lead to a general improvement in the
quality of the fit, and to some significant changes in partons and
predictions, particularly at small $x$. We have not considered
NNLO at all here, but will produce detailed results in a
forthcoming paper \cite{MRSTNNLO}.

In summary, in this paper we have used all deep inelastic and
hadron collider data available in order to obtain the most
accurate and precise determination of the NLO parton distributions
currently in existence, and have also determined the value of the
strong coupling constant $\alpha_S(M_Z^2)$ with tight constraints.
This enables us to probe the success of the conventional NLO
perturbative QCD framework in describing hadronic collider
physics, and we find that overall it is still working well. This
then provides us with the necessary starting point for predicting
and explaining new physics coming from present and future particle
colliders.

The FORTRAN code for the four NLO parton sets mentioned in Table~1
can be found at http://durpdg.dur.ac.uk/hepdata/mrs

\section*{Acknowledgments}

We would like to thank Levan Babukhadia, Iain Bertram, Anwar
Bhati, Arie Bodek, John Collins, Mandy Cooper-Sarkar, Brenna
Flaugher, Brian Foster, Nigel Glover, Eram Rizvi, Olaf Ruske,
Neils Tuning, Andreas Vogt, Rainer Wallny, Un-ki Yang and Rik
Yoshida for useful discussions and information concerning the
data.  RST would like to thank the Royal Society for the award of
a University Research Fellowship. This work was supported in part
by the EU Fourth Framework Programme ``Training and Mobility of
Researchers'', Network ``Quantum Chromodynamics and the Deep
Structure of Elementary Particles'', contract FMRX-CT98-0194(DG 12
- MIHT).

\section*{Appendix}

We have investigated the effects of fitting the HERA data 
\cite{H1A,H1B,H1C,ZEUS} taking into account the systematic errors 
in a consistent fashion. In order to do this we have adopted the same 
procedure as for the CDF jet data, i.e. in order to obtain the $\chi^2$ we 
use (\ref{eq:sys1}), which is also the procedure usually adopted by the H1
collaboration. As before we limit the size of each of the $s_k$ to be 
$\leq 1$, though again this has little effect.

Let us discuss the fit to the ZEUS data first, since this is particularly 
simple. If we take our default fit and compare to the ZEUS data with 
uncorrelated errors only, we obtain a $\chi^2$ of 378 for the 242 points in 
\cite{ZEUS}. If we keep the theory fixed and let the $s_k$ for the 
sources of correlated errors vary then $\chi^2$ lowers to 331. The way in 
which 
it does this is very simple -- the majority of the data normalizes down to its
minimum of $98\%$, while that below $Q^2 = 27~{\rm GeV}^2$ takes its 
normalization down the further available $1\%$. This seems to be essentially 
in order to bring the data into line with the H1 normalization 
(which matches well with that of the NMC data). All further changes are 
a very minor perturbation to this. Those error sources which could alter 
the shape, such 
as the positron energy scale and hadronic energy flow labeled type B, 
play no part, 
presumably because they would cause dramatic alterations in a few bins which
would only lead to a deterioration of the fit. 

The effect for H1 data is a little more complicated. The default fit using 
uncorrelated errors alone gives a $\chi^2$ of 485 for the 400 points. 
If we let the $s_k$ for the 
sources of correlated errors vary, then $\chi^2$ lowers to 381. 
This comes from 
3 sources. Some of the low $x$ and $Q^2$ points move up $\sim 2\%$, coming 
closer to the ZEUS data. Some of the highest $x$ points use their large 
correlated errors to move up $5-10\%$ since they clearly fell below theory
(and the extrapolation of BCDMS data). Finally, the biggest improvement 
comes from the region $0.001 < x < 0.08$, where the low $Q^2$ points move up
by up to $3\%$, and the high $Q^2$ points minimize their normalization to
move down by $2-3\%$. These effects combine to flatten the slope with $Q^2$ at
fixed $x$ and partially reconcile the data with the failure of the theory 
to have a large enough $dF_2(x,Q^2)/d\ln Q^2$ in this range of $x$.    
      
Thus the data shift from their central values to partially account for either 
the incompatibility between H1 and ZEUS data, or between data and theory. If 
the data are then moved from their central values to those imposed by the 
correlated systematic errors and the fit redone there is to all intents and 
purposes no change in either $\alpha_S(M_Z^2)$ or the partons,
i.e the iteration essentially converges at the first step. 
To be precise, the value of $\alpha_S(M_Z^2)$ for the best fit moves by less 
than $0.0003$, the quark distributions change by less than $0.5\%$ at all $x$ 
and $Q^2$, and the gluon distribution changes by a maximum of $1.5\%$ at 
$Q^2 \sim 2~{\rm GeV}^2$, lowering to $1\%$ at $Q^2 = 
5~{\rm GeV}^2$ and $0.5\%$ at $Q^2 > 20~{\rm GeV}^2$.\footnote{The percentage 
change in the gluon distribution
is large at low $Q^2$ in the precise region of $x$ where the gluon 
distribution becomes negative, but this is just due to the change in sign 
of the gluon here, and the absolute change is extremely small.}    
Hence,
the correlated systematic errors simply allow the data to readjust themselves 
to best match the possible theory and other data sets, but rather
surprisingly the
parameters in the theory do not alter at all significantly to rematch the
altered data. We summarize the major effects. First, the normalization 
change of the ZEUS measurements, and of low $Q^2$, 
low $x$ H1 measurements brings the data sets closer together, but does not 
change the best fit, which is a still a compromise between them. 
Second,the raising of the high $x$ H1 data 
improves its description, but these data carry virtually no weight compared 
to NMC, BCDMS and SLAC data. Finally, the flattening of the intermediate $x$ 
data in $Q^2$ again improves the fit quality, but $dF_2(x,Q^2)/d\ln Q^2$ is 
still too flat
for these data even after the alteration, and also for NMC data, so the pull 
on the fit is only partly diminished.

In fact we also notice that the increments in $\chi^2$ between different fits 
are very similar when the correlated systematic errors are used in full to when
the simple addition in quadrature of statistical and systematic errors is used.
This can be seen by examining the fits for $\alpha_S(M_Z^2)=0.116$ to
$\alpha_S(M_Z^2)=0.122$ using the iterative procedure described above.
We find that although the base points of the curves for HERA data in the 
bottom right plot in Fig. 16 move (significantly for ZEUS), the shape and
scale of the curves is extremely similar to those shown 
-- there is a very slight 
tendency for the low $\alpha_S(M_Z^2)$ $\chi^2$ to be lower and the high 
$\alpha_s(M_Z^2)$ $\chi^2$ to be higher, but only by a couple of units.
The refit in the second stage of the iteration results in the $\chi^2$ for 
other data sets changing, but only by a couple of points each, and in such 
a way that 
the total change for non-HERA data is only a couple of points in total. 
Hence, the curve for the total $\chi^2$ in the top left of Fig. 16 is 
changed by at most 3--4 units at each $\alpha_S(M_Z^2)$, except for the 
common shift, and all conclusions on parton distributions for the best fit
and uncertainty of $\alpha_S(M_Z^2)$ are unaltered.          
We have also examined more significant changes in theory curves 
by comparing to a theoretical model designed to work 
better at small $x$. In this case the fit to H1 data gives a better 
description of $dF_2(x,Q^2)/d\ln Q^2$, 
with a $\chi^2$ of 420, if uncorrelated errors alone are 
used. When correlated errors are allowed to contribute, no large correction is 
needed for intermediate $x$ so the improvement is of $420-81 =339$ compared to
$485-104 = 381$ above. In both cases the final $\chi^2$ values 
are within a 3--4 units
of that obtained from the simplistic procedure. For ZEUS data the model gives
a $\chi^2$ value 18 worse for quadrature, and 23 worse for the full treatment 
of errors, and in this case the increment when statistical errors alone are 
used is much the same. 
    
Hence, the freedom of the data to move in the direction of preferred theory
or other data when correlated systematic errors are properly accounted for 
lessens the pull on a fit compared to the use of uncorrelated errors alone,
but for the present HERA data this seems to have a rather similar effect to 
the lessening of the pull obtained by adding statistical and systematic errors 
in quadrature. The relative success of this approximation 
is presumably due to the fact that for the vast majority 
of the data points the uncorrelated error is easily dominant, and correlated 
shifts in the data are rather smaller than the uncorrelated errors 
of most points. This is in contrast to the Tevatron jet data where, as we
see in Fig. 12, the correlated shift can be an order of magnitude greater
than the uncorrelated error of some points.          
Therefore, in our determination of the best fits, and the
variations about these, we use the simple prescription for errors for 
the HERA data, since it
does not lead to unnecessary complications, and does not change any results
to any significant degree, as quantified above.

\newpage

\begin{figure}[H]
\begin{center}
\epsfig{figure=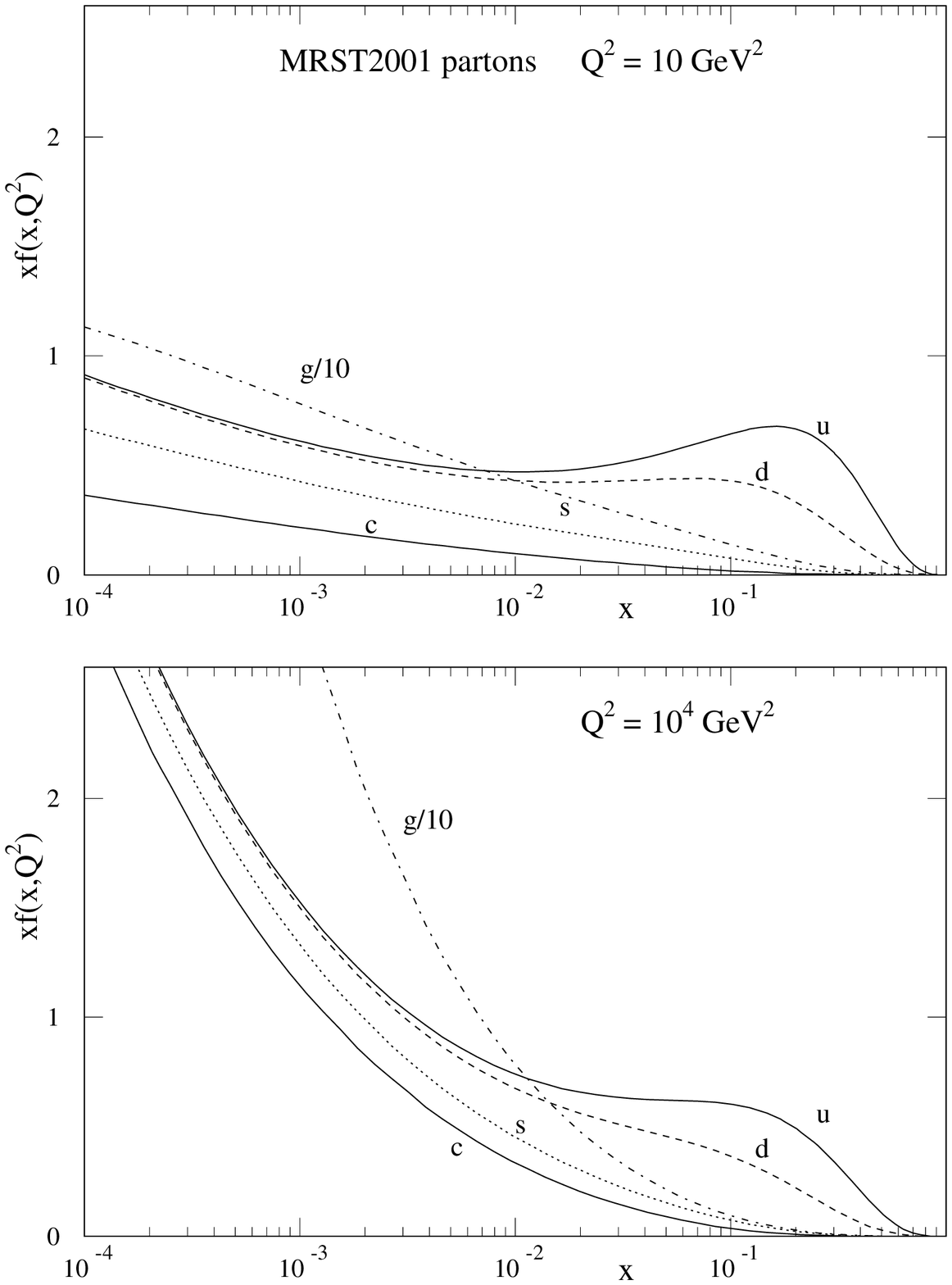,height=20cm}
\end{center}
\caption{MRST2001 partons at $Q^2=10~{\rm GeV}^2$ and $Q^2 =
10^4~{\rm GeV}^2$.} \label{fig:Fig1}
\end{figure}

\begin{figure}[H]
\begin{center}
\epsfig{figure=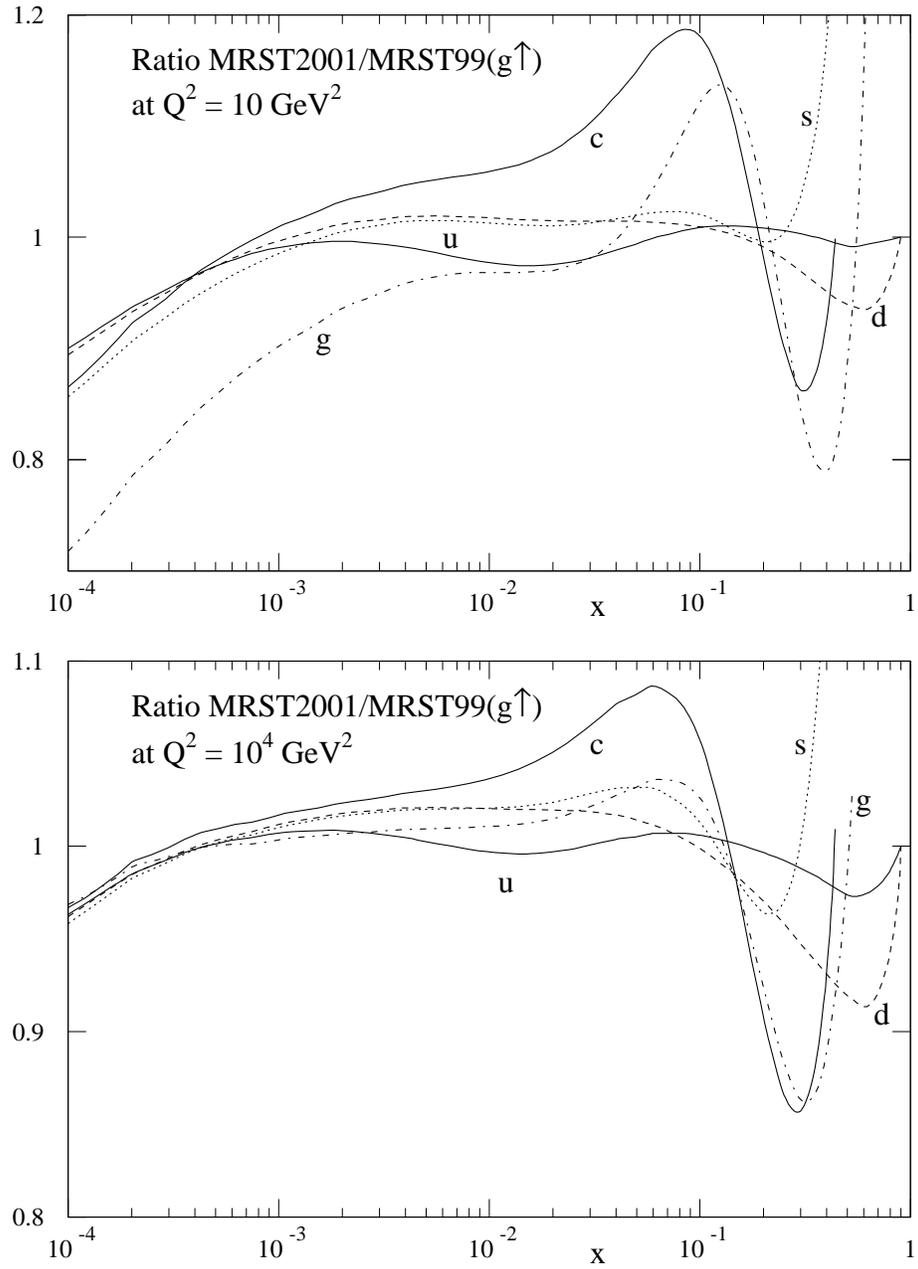,height=20cm}
\end{center}
\caption{Comparison of the MRST2001 partons with those of
MRST99$(g\!\uparrow)$ \cite{MRST99} at $Q^2=10~{\rm GeV}^2$ and
$Q^2=10^4~{\rm GeV}^2$.} \label{fig:Fig2}
\end{figure}

\begin{figure}[H]
\begin{center}
\epsfig{figure=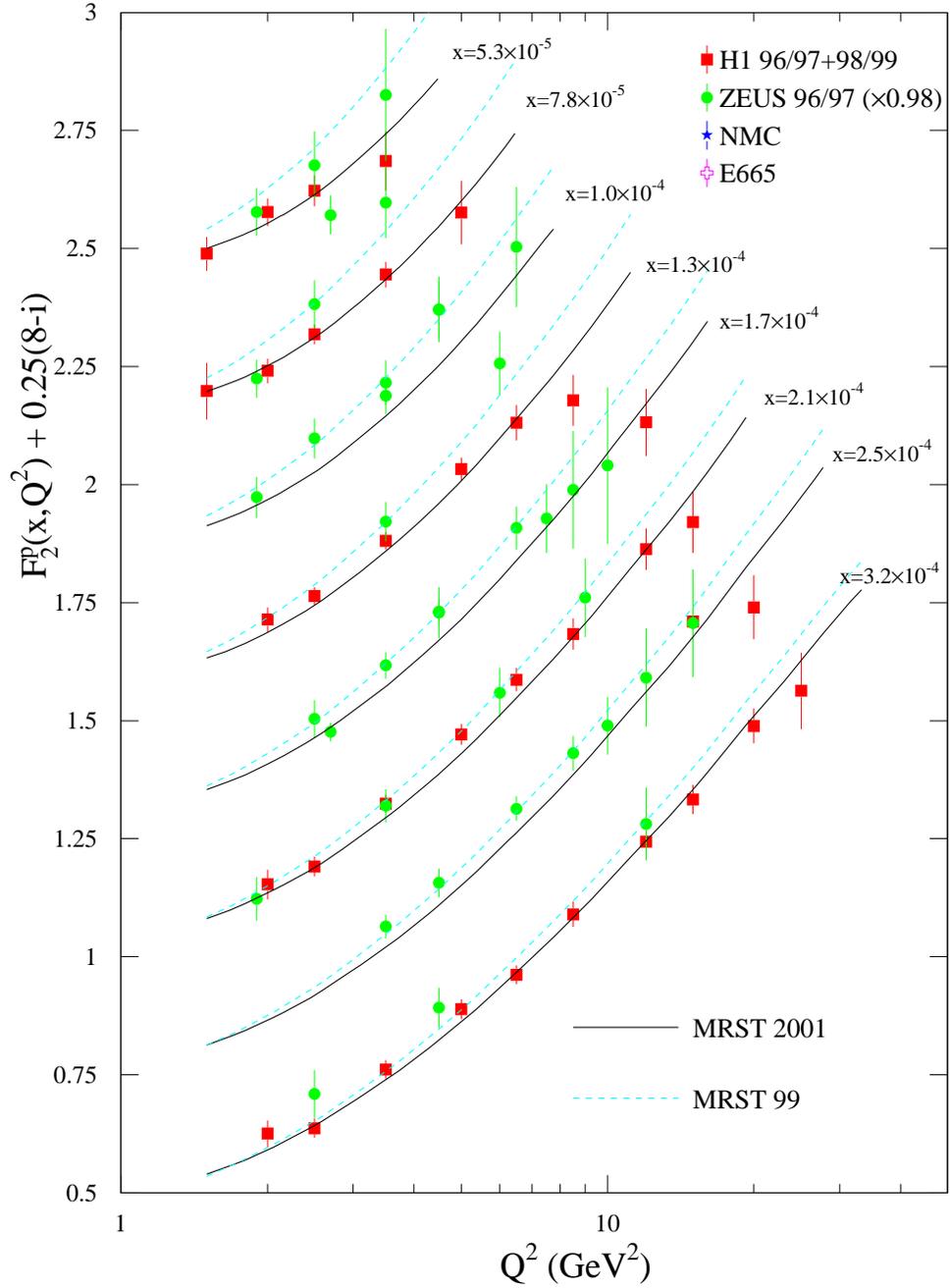,height=20cm}
\end{center}
\caption{Comparison of the MRST2001 prediction of $F_2(x,Q^2)$
with data and with MRST99 for $x= 0.00005-0.00032$.
The error bars show statistical and systematic errors added in quadrature.}
\label{fig:Fig3}
\end{figure}

\begin{figure}[H]
\begin{center}
\epsfig{figure=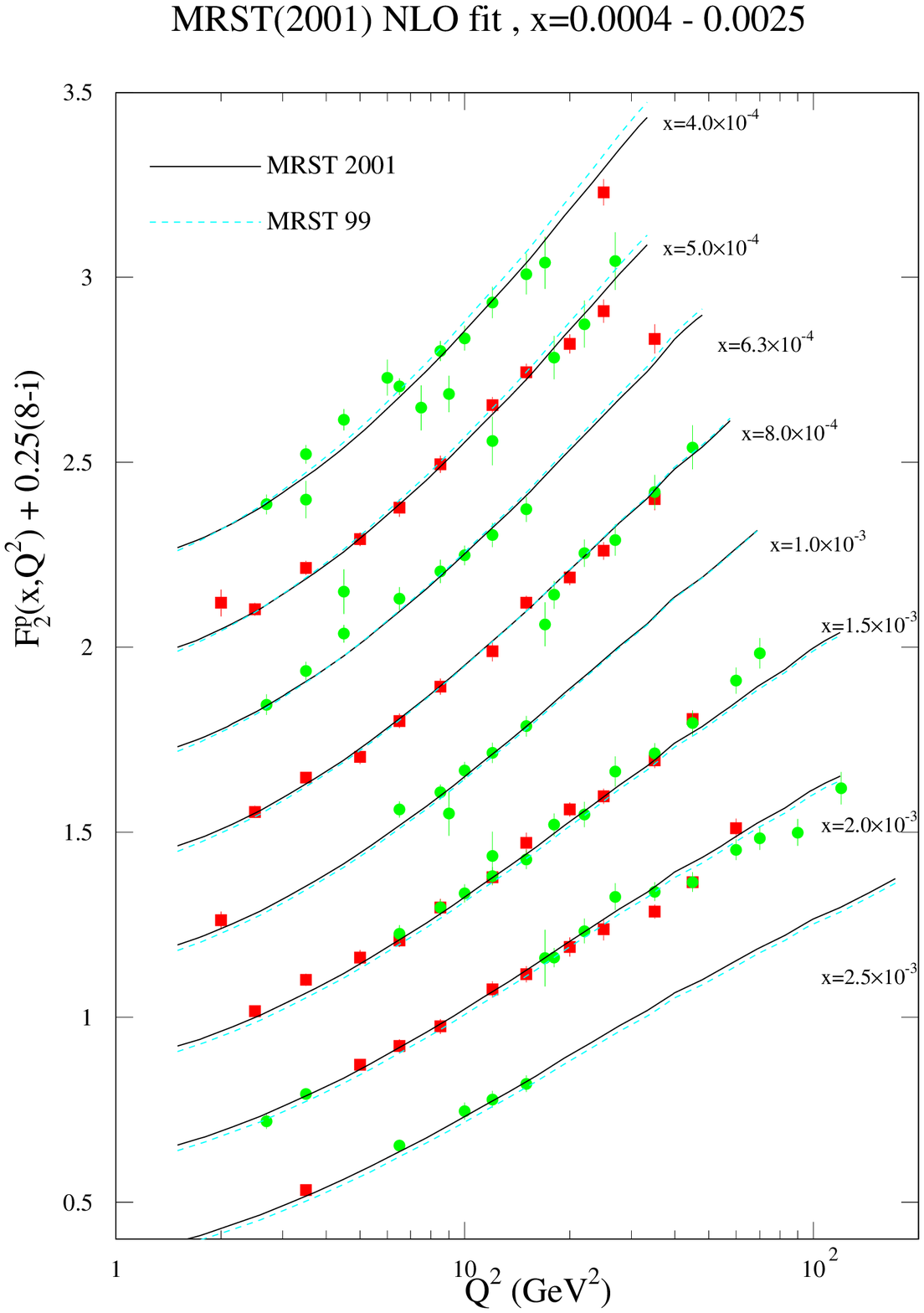,height=20cm}
\end{center}
\caption{Comparison of the MRST2001 prediction of $F_2(x,Q^2)$
with data and with of MRST99 for $x= 0.0004-0.0025$.
The data points are as indicated in Fig. 3.}
\label{fig:Fig4}
\end{figure}

\begin{figure}[H]
\begin{center}
\epsfig{figure=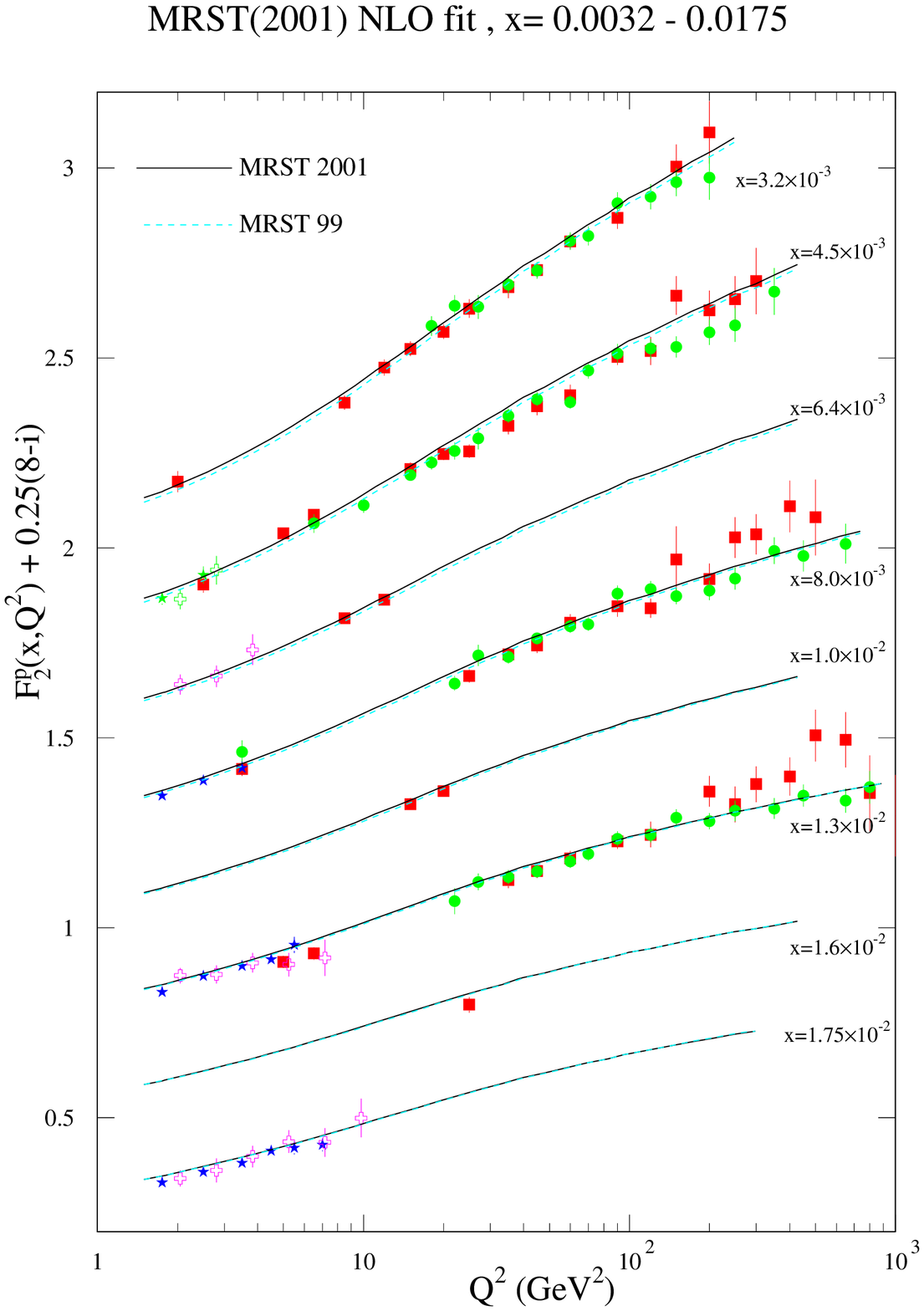,height=20cm}
\end{center}
\caption{Comparison of the MRST2001 prediction of $F_2(x,Q^2)$
with data and with MRST99 for $x= 0.0032-0.0175$.
The data points are as indicated in Fig. 3.}
\label{fig:Fig5}
\end{figure}

\begin{figure}[H]
\begin{center}
\epsfig{figure=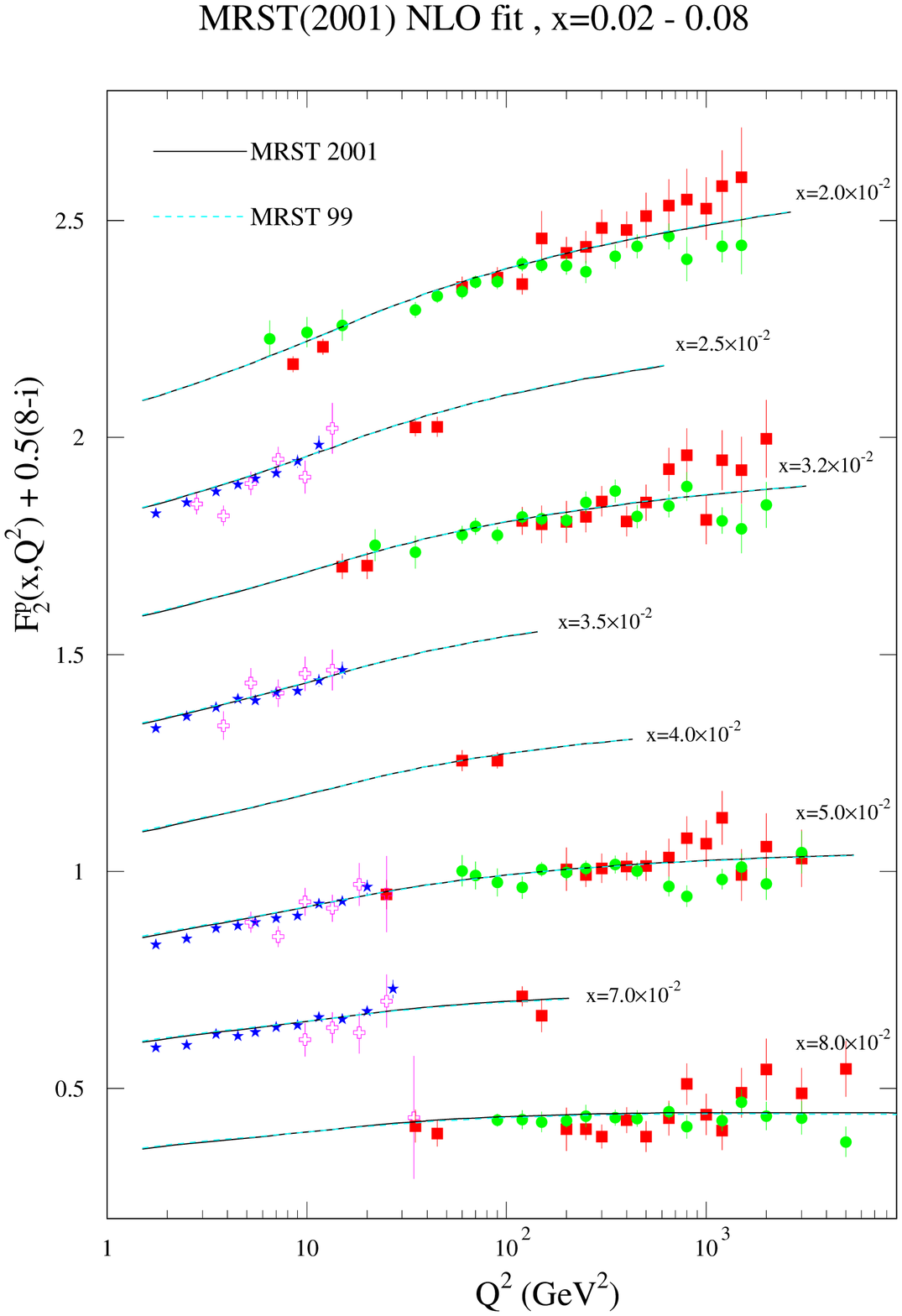,height=20cm}
\end{center}
\caption{Comparison of the MRST2001 prediction of $F_2(x,Q^2)$
with data and with MRST99 for $x= 0.02-0.08$.
The data points are as indicated in Fig. 3.}
\label{fig:Fig6}  
\end{figure}


\begin{figure}[H]
\begin{center}
\epsfig{figure=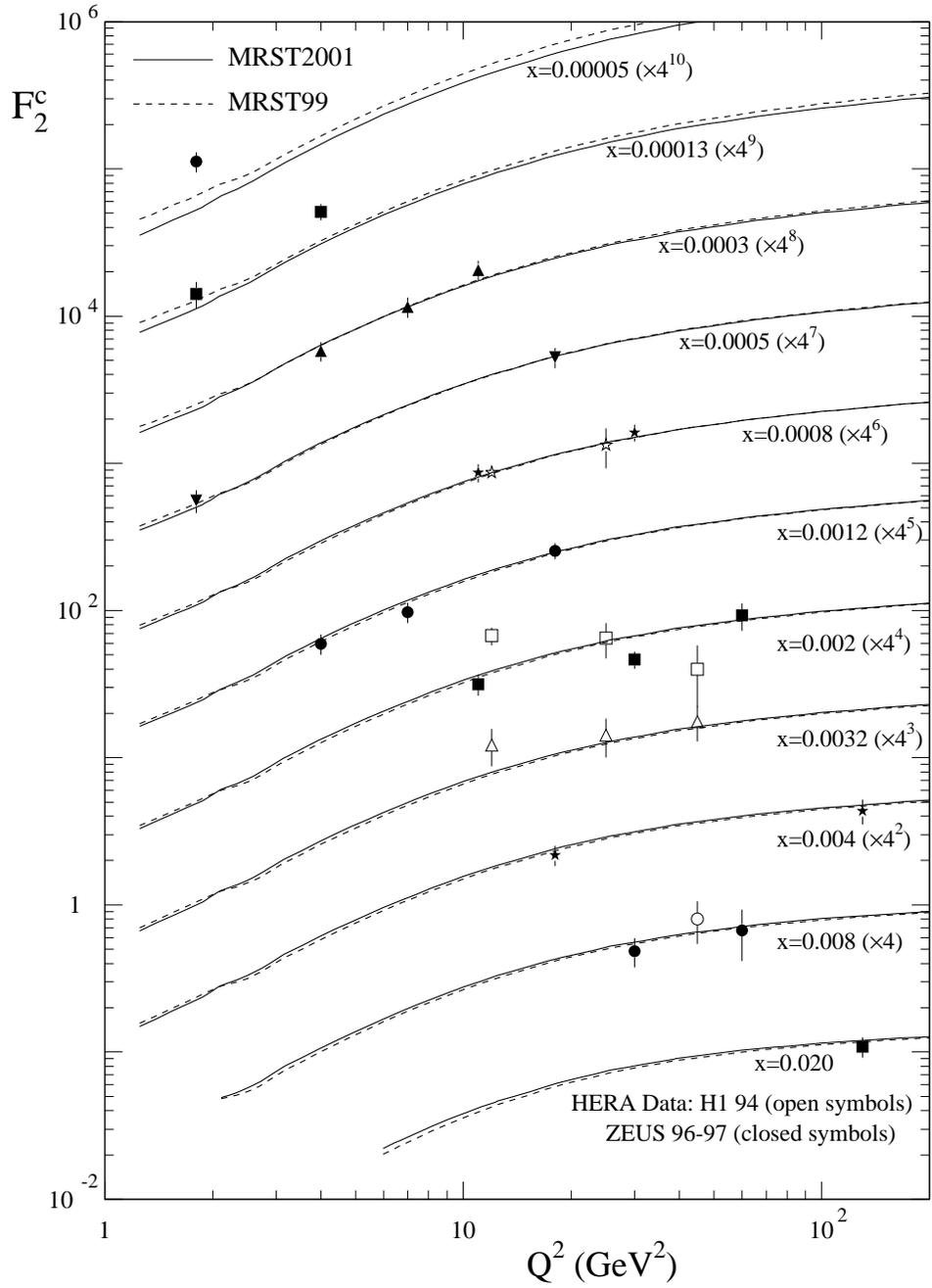,height=20cm}
\end{center}
\caption{Comparison of the MRST2001 prediction of $F^c_2(x,Q^2)$
with data \cite{ZEUSc,H1charm}.} \label{fig:Fig7}
\end{figure}

\begin{figure}[H]
\begin{center}
\epsfig{figure=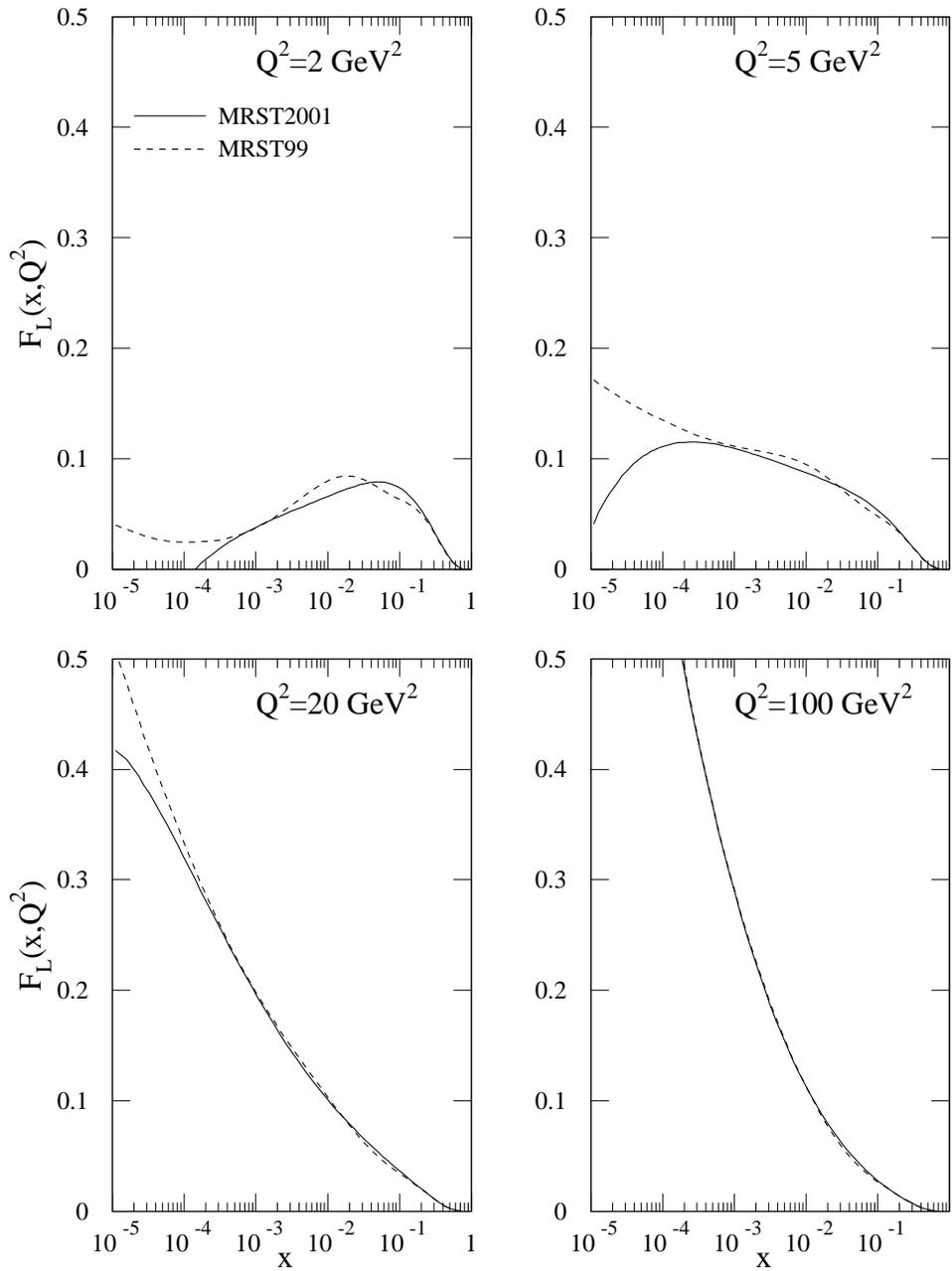,height=20cm}
\end{center}
\caption{Comparison of the MRST2001 prediction of $F_L(x,Q^2)$
with that of MRST99 for various values of $Q^2$.} \label{fig:Fig8}
\end{figure}

\begin{figure}[H]
\begin{center}
\epsfig{figure=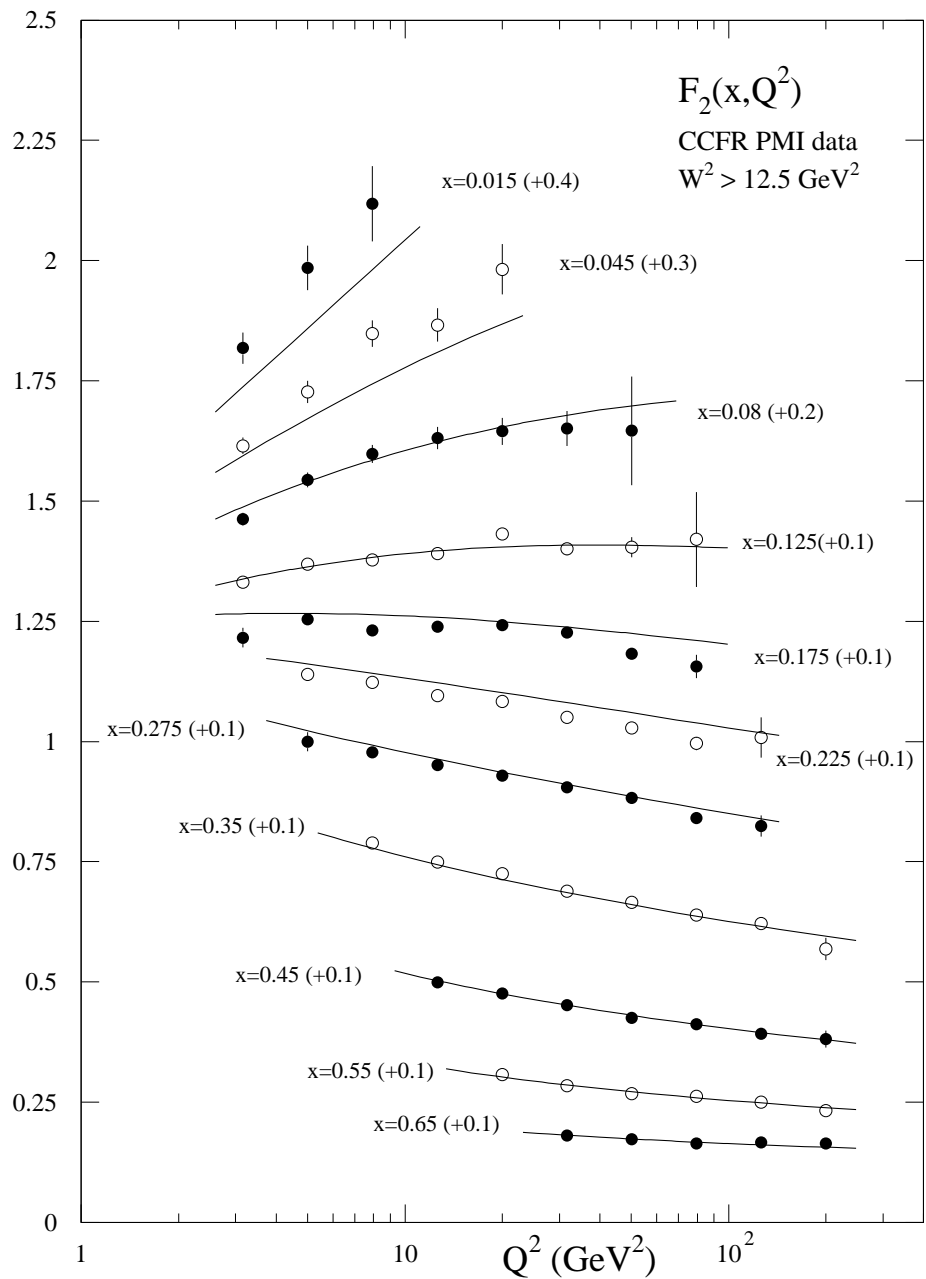,height=20cm}
\end{center}
\caption{Quality of the MRST2001 fit to the CCFR $F_2^{\nu(\bar
\nu)N}(x,Q^2)$ PMI data \cite{CCFR}.} \label{fig:Fig9}
\end{figure}

\begin{figure}[H]
\begin{center}
\epsfig{figure=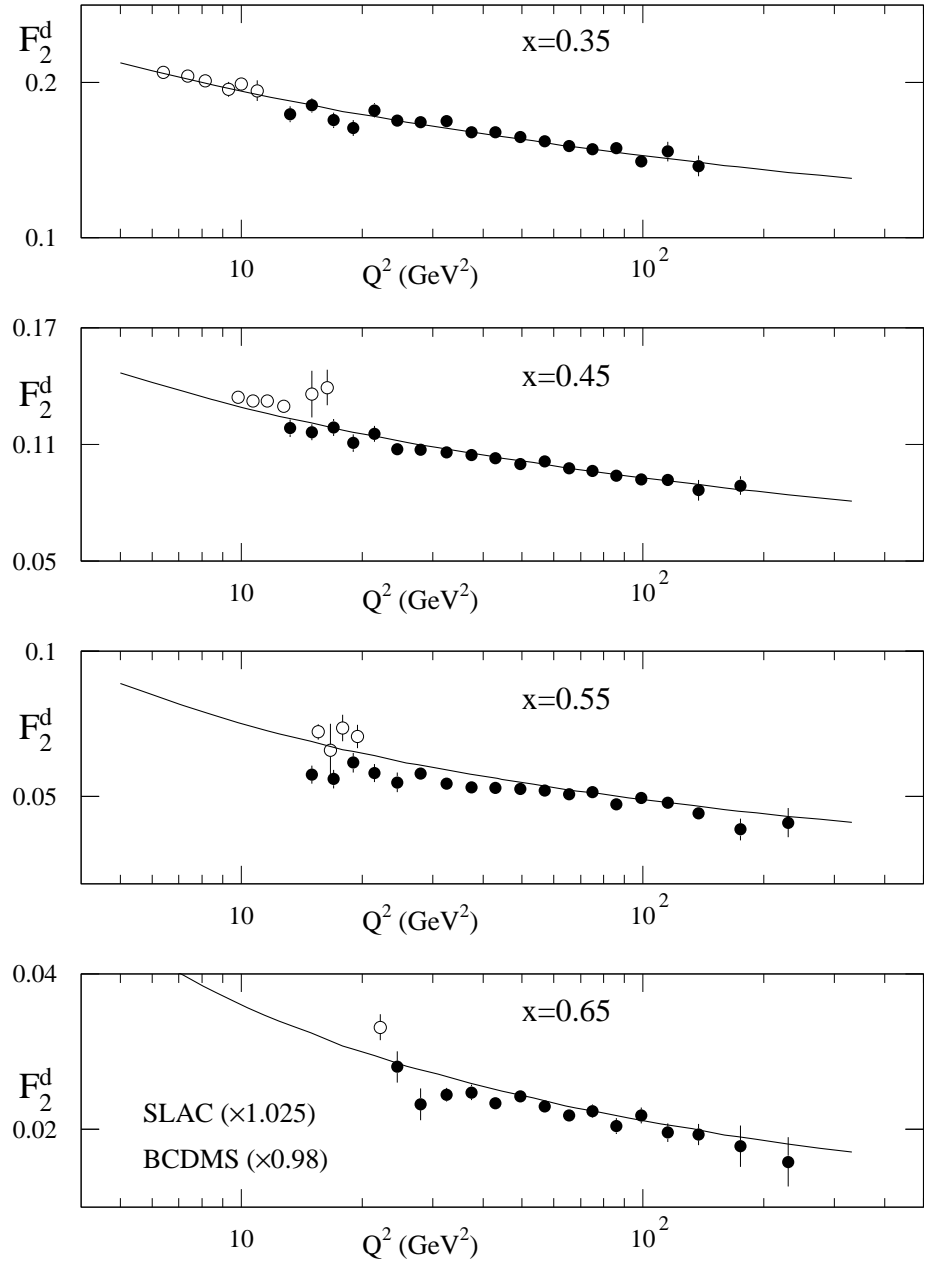,height=20cm}
\end{center}
\caption{Quality of the MRST2001 fit to the SLAC and BCDMS
deuterium structure function data \cite{SLAC,BCDMSd} at high $x$.}
\label{fig:Fig10}
\end{figure}

\begin{figure}[H]
\begin{center}
\epsfig{figure=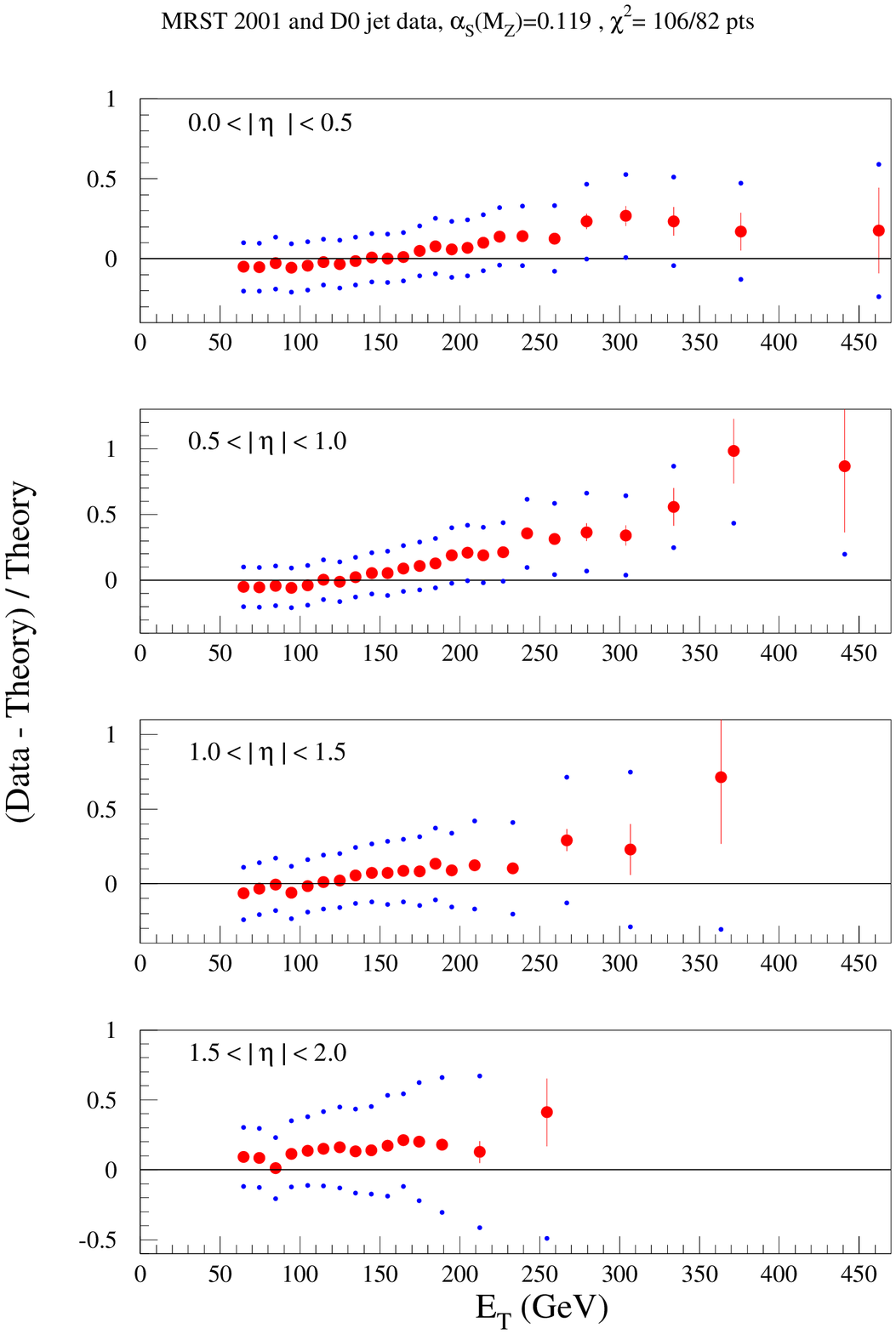,height=20cm}
\end{center}
\caption{Quality of the MRST2001 fit to the D0 high $E_T$ jet data
in different $\eta$ bins \cite{D0}. The band shows the 
allowed shift from the central value for each point obtained by adding the 
correlated errors in quadrature.} \label{fig:Fig11}
\end{figure}

\begin{figure}[H]
\begin{center}
\epsfig{figure=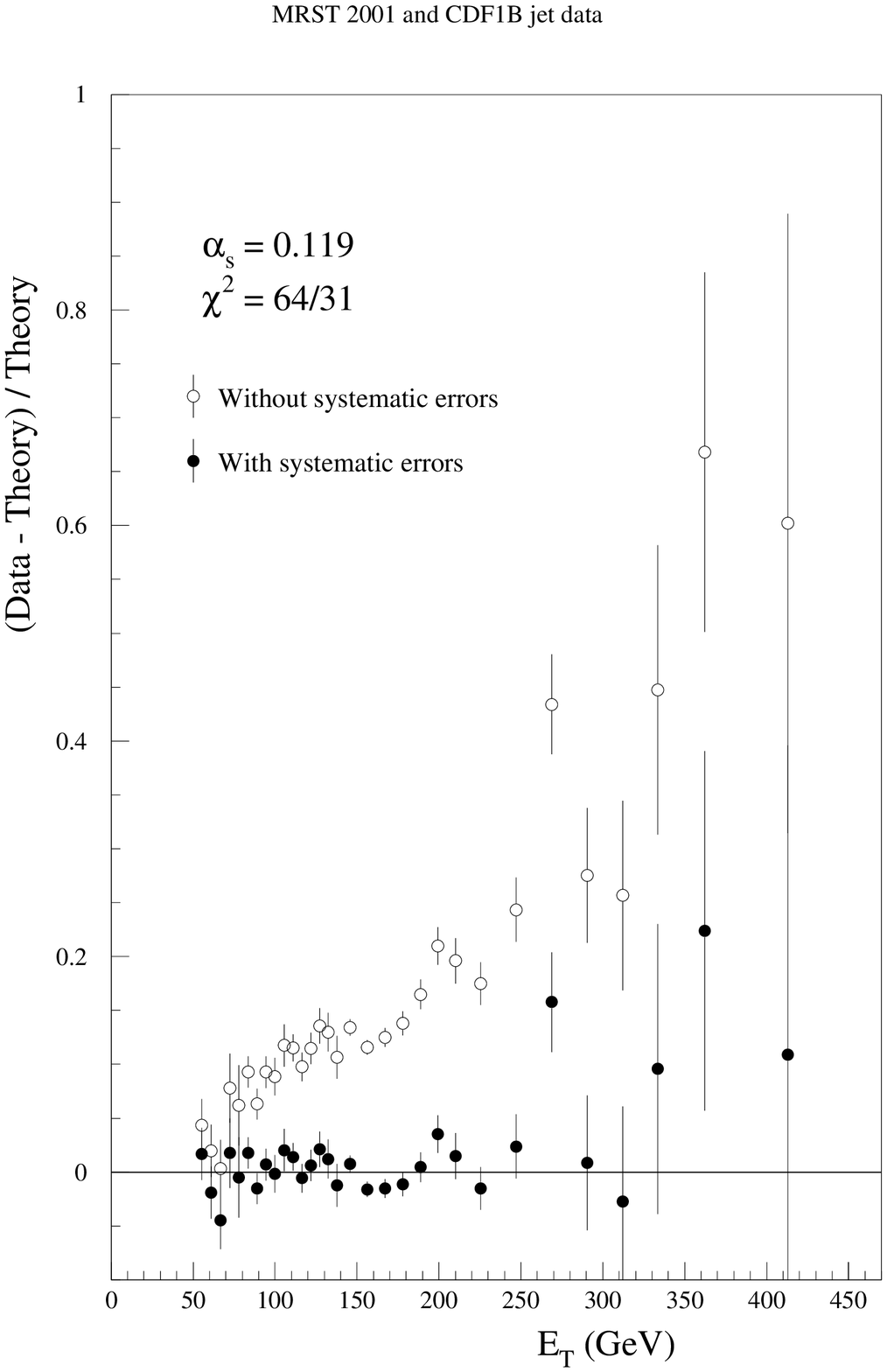,height=20cm}
\end{center}
\caption{Quality of the MRST2001 fit to the CDF1B high $E_T$ jet
data \cite{CDF}. The open points are before correlated systematic
errors have been considered, while the solid points are after the
correlated errors have allowed the data-theory comparison to move
at some cost to the total $\chi^2$ (shown on the plot).}
\label{fig:Fig12}
\end{figure}

\begin{figure}[H]
\begin{center}
\epsfig{figure=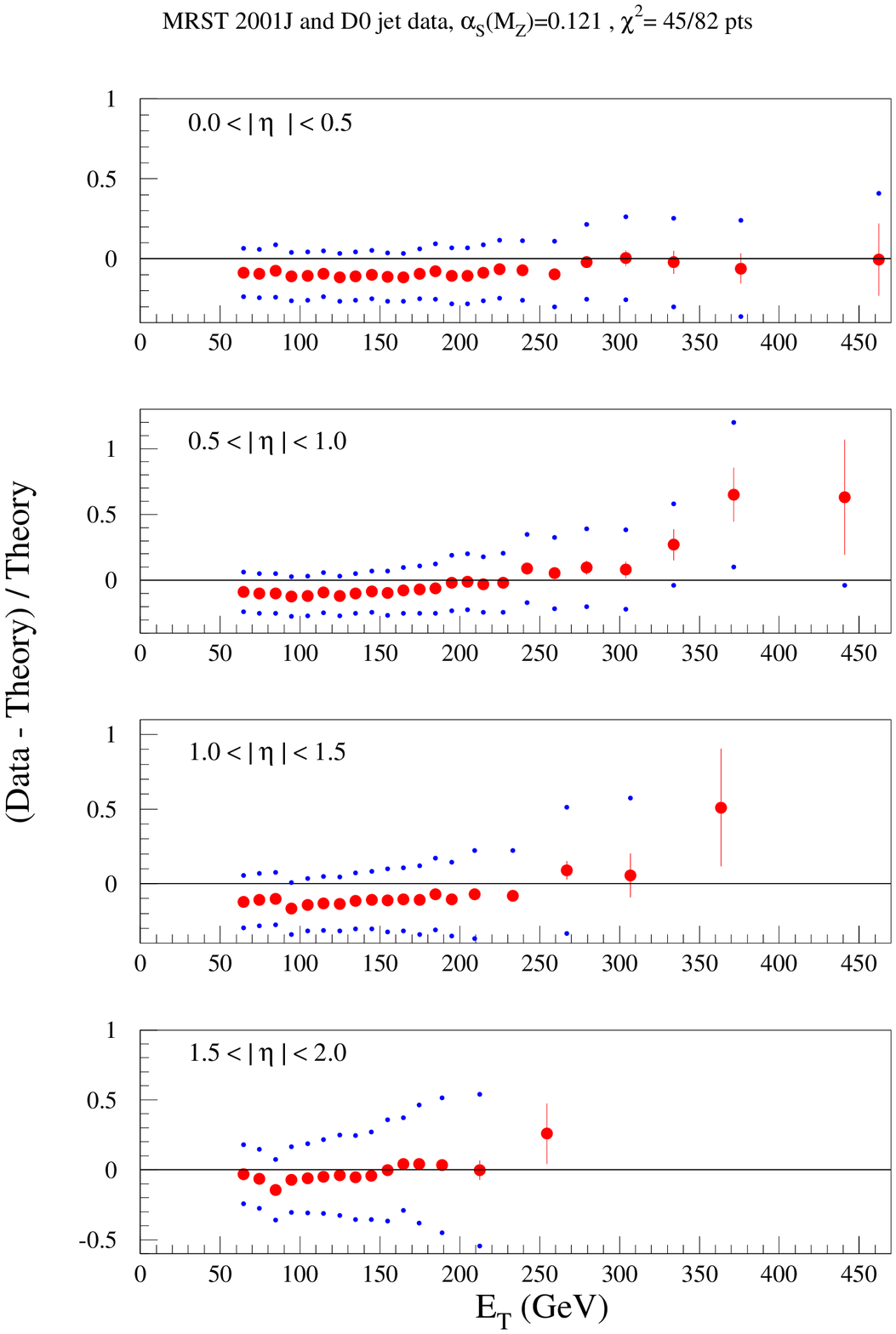,height=20cm}
\end{center}
\caption{Quality of the MRST2001J fit to the D0 high $E_T$ jet
data \cite{D0} in different $\eta$ bins. The band shows the 
allowed shift from the central value for each point obtained by adding the 
correlated errors in quadrature.} \label{fig:Fig13}
\end{figure}

\begin{figure}[H]
\begin{center}
\epsfig{figure=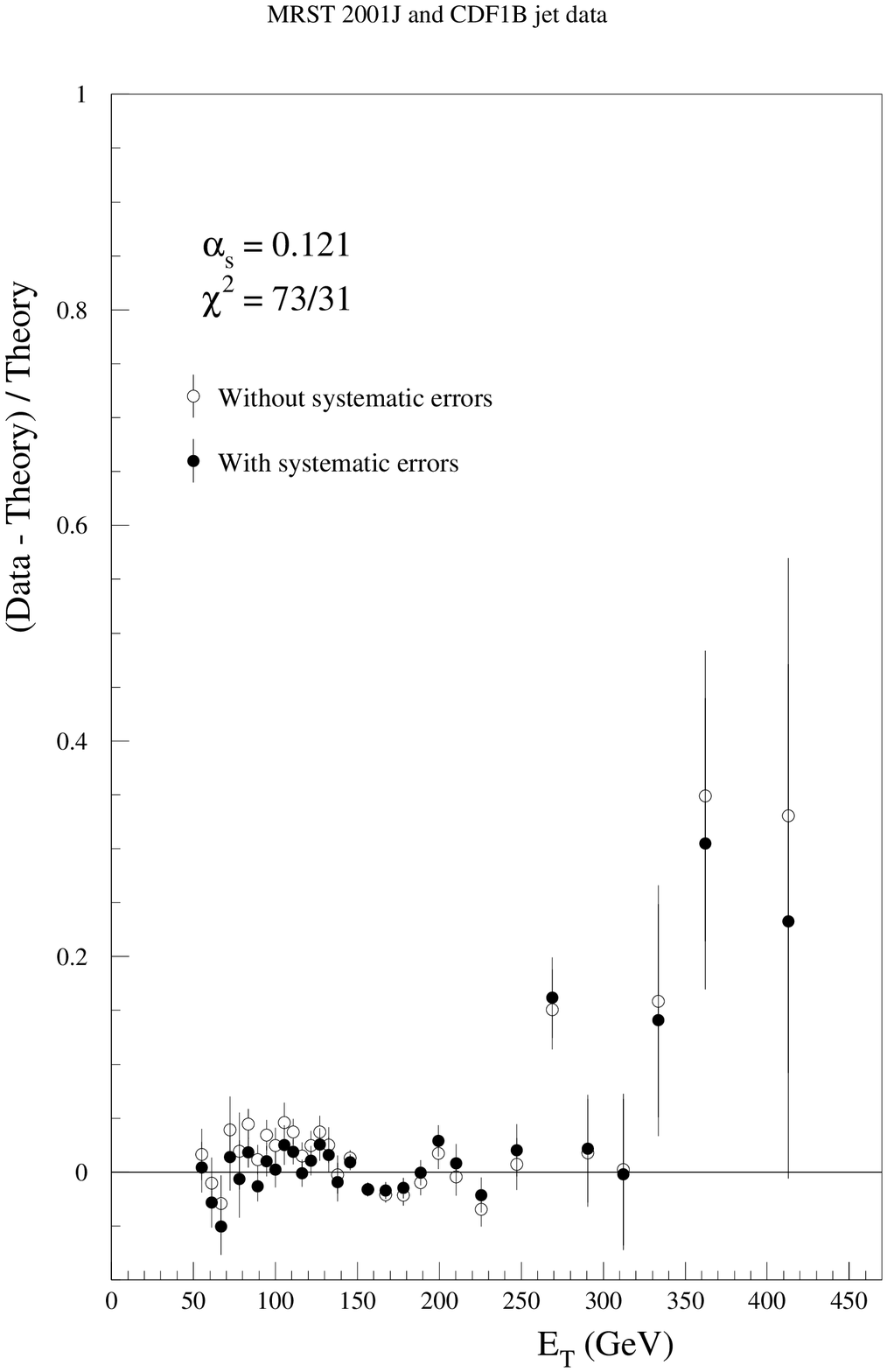,height=20cm}
\end{center}
\caption{Quality of the MRST2001J fit to the CDF1B high $E_T$ jet
data \cite{CDF}. The open points are before correlated systematic
errors have been considered, while the solid points are after the
correlated errors have allowed the data-theory comparison to move
at some cost to total $\chi^2$ (shown on the plot).}
\label{fig:Fig14}
\end{figure}

\begin{figure}[H]
\begin{center}
\epsfig{figure=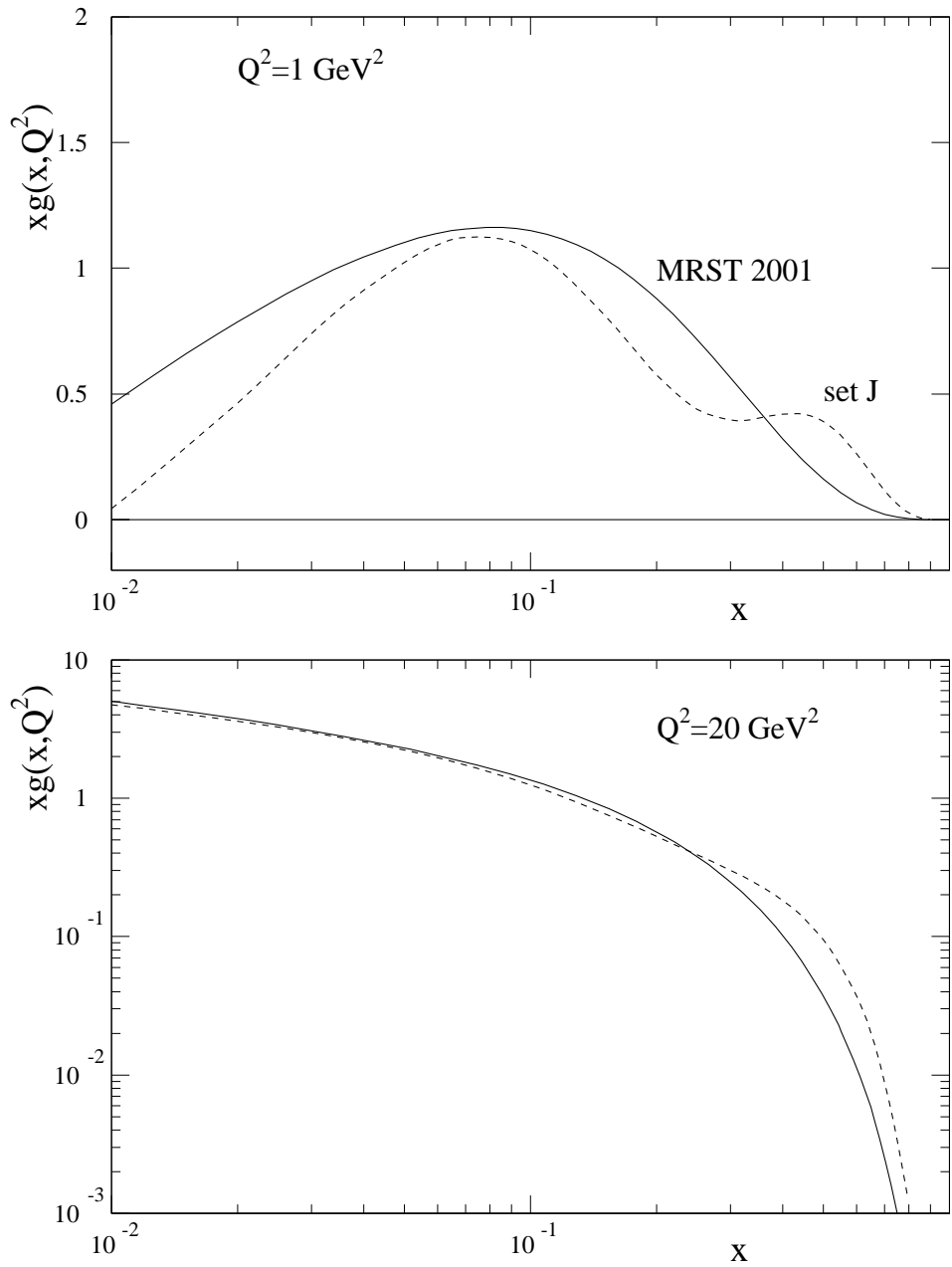,height=20cm}
\end{center}
\caption{Comparison of the MRST2001J gluon with the MRST2001 gluon
at high $x$ and $Q^2= 1~{\rm GeV}^2$ and  $Q^2= 20~{\rm GeV}^2$.}
\label{fig:Fig15}
\end{figure}

\begin{figure}[H]
\begin{center}
\epsfig{figure=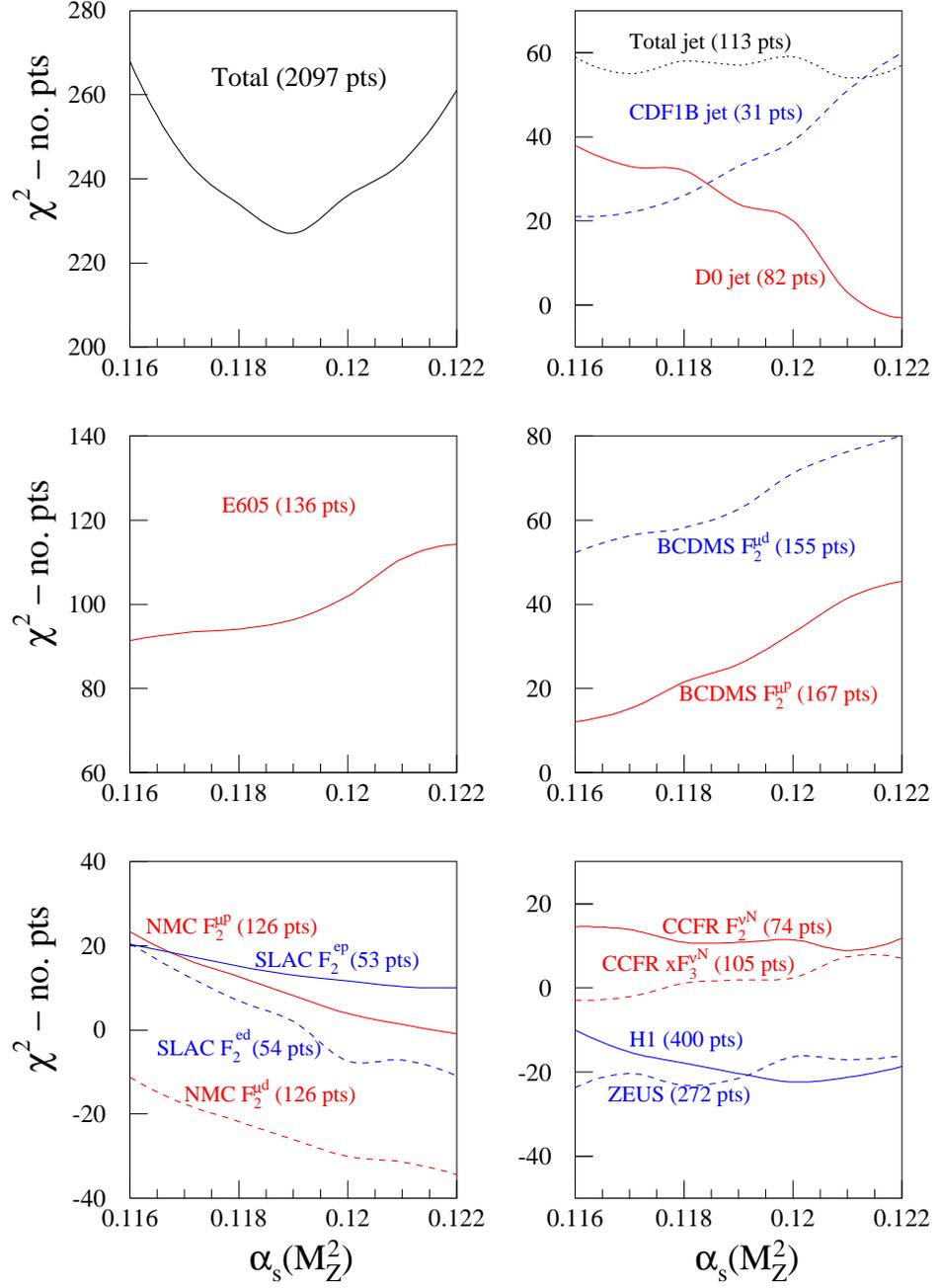,height=20cm}
\end{center}
\caption{The quality of the fit to the individual data sets
included in the global analysis, shown together with the grand
total $\chi^2$, as a function of $\alpha_S(M_Z^2)$. In total there are 
2097 data points, 23 parameters for the parton distributions and 5 free 
normalizations for data sets.}
\label{fig:Fig16}
\end{figure}

\end{document}